\documentclass[a4paper,fleqn,usenatbib]{mnras}

\usepackage{newtxtext,newtxmath}

\usepackage[T1]{fontenc}
\usepackage{ae,aecompl}

\usepackage{graphicx,psfig}	
\usepackage{amsmath}	
\usepackage{amssymb}	
\usepackage{graphics}

\usepackage{longtable}
\usepackage[font=small,labelfont=bf,margin=\parindent,tableposition=top]{caption}
\usepackage[usenames,dvipsnames,svgnames,table]{xcolor}

\usepackage{hyperref}
 \definecolor{arancio}{rgb}{1,0.5,0}
 \definecolor{viola}{rgb}{0.7,0,1}
 \definecolor{verde}{rgb}{0.2,0.7,0.7}
\def \pmn  {{PMN~J0948+0022}} 
\def \sbs  {{SBS~0846+513}} 
\def \pks {{PKS~1502+036}} 
\newcommand\kms{\ifmmode {\rm km\ s}^{-1} \else km s$^{-1}$\fi} 
\newcommand\FWHM{\ifmmode {\rm FWHM} \else ${\rm FWHM}$\fi}
\newcommand\Lsun{\ifmmode L_{\odot} \else $L_{\odot}$\fi} 
\newcommand\Hbeta{\ifmmode {\rm H}\beta 
 \else H$\beta$\fi} 

\def \ATel {ATel} 
\def \apj {ApJ}
\def \apjl {ApJL}

\def \aap {A\&A}

\def \mnras {MNRAS}

\title[$\gamma$-NLS1s with CTA]{Prospects 
for gamma-ray observations of narrow-line Seyfert 1 galaxies with the Cherenkov Telescope Array}

\author[P. Romano et al.]{
P.\ Romano,$^{1}$\thanks{E-mail: patrizia.romano@inaf.it}
S.\ Vercellone,$^{1}$
L.\ Foschini,$^{1}$
F.\ Tavecchio,$^{1}$
M.\ Landoni$^{1}$ 
\newauthor  and J.\ Kn{\"o}dlseder$^{2}$ \\
$^{1}$INAF, Osservatorio Astronomico di Brera, Via E.\ Bianchi 46, I-23807, Merate, Italy\\
$^{2}$Institut de Recherche en Astrophysique et Plan\'etologie, 9 avenue Colonel-Roche, 31028, Toulouse Cedex 4, France 
}

\date{Accepted 2018 September 6. Received 2018 August 27; in original form 2018 July 25}

\pubyear{2018}

\begin{document}
\label{firstpage}
\pagerange{\pageref{firstpage}--\pageref{lastpage}}
\maketitle

\begin{abstract}
Gamma-ray emitting narrow-line Seyfert 1 ($\gamma$-NLSy1) galaxies are thought to  
  harbour relatively low-mass black holes (10$^6$--10$^8$\,M$_{\sun}$) accreting 
  close to the Eddington limit. They show characteristics similar 
  to those of blazars, such as flux and spectral variability in the gamma-ray energy
  band and radio properties which point toward the presence of a relativistic jet. 
  These characteristics make them an intriguing class of sources to be investigated 
  with the Cherenkov Telescope Array (CTA), the next-generation ground-based
  gamma-ray observatory. 
  We present our extensive set of simulations of all currently known $\gamma$-ray
  emitters identified as NLS1s (20 sources),  
  investigating their detections and spectral properties, taking into 
  account the effect of both the extra-galactic background light in the
  propagation of gamma-rays and intrinsic absorption components. 
  We find that the prospects for observations of  $\gamma$-NLSy1 with CTA are promising. 
  In particular, the brightest sources of our sample, SBS 0846+513,
  PMN J0948+0022, and PKS 1502+036 can be detected during high/flaring states, the former two 
  even in the case in which the emission occurs within the highly opaque central regions, 
  which prevent $\gamma$ rays above few tens of GeV to escape. In this case the low-energy
  threshold of CTA will play a key role. 
  If, on the other hand, high-energy emission occurs outside the broad line region, 
  we can detect the sources up to several hundreds of GeV--depending on the intrinsic
  shape of the emitted spectrum. 
  Therefore, CTA observations will provide valuable information on the physical conditions
  and emission properties of their jets.
\end{abstract}




 	 \section{Introduction}  \label{romano_nls1:intro}

\setcounter{table}{0} 
 \begin{table*} 
\begin{center}
\small	
 \tabcolsep 3pt  
 \caption{Sample of $\gamma$-NLS1 and spectral parameters adopted for the simulations.}
  \label{romano_nls1:table:sample} 
  \begin{tabular}{ll rr c  c crrc c } 
 \hline 

 \noalign{\smallskip} 
    Source Name&  Common Name                                                                  & RA     & Decl       	 & $z$ & Model &                   &                      &                &                  & Ref. \\
                                            &                                                                        &           &  	         &         &      PL     &   $K_{0}$              &   $E_{0}$         &  $\Gamma$                 &    &   \\    
 	                                    &                                                                        &           &  	          &        &      LP     &                            &                       &  $\alpha$                 &  $\beta$  &   \\    
 	                                    &                                                                        &           &  	          &        &      BKPL  &                           &  $E_{b}$          &  $\Gamma_1$        & $\Gamma_2$    &      \\    
                                            &                                                                        &  (deg) & (deg)	         &         &        & (ph\,cm$^{-2}$\,MeV$^{-1}$\,s$^{-1}$)    & (MeV)  &        &         \\
 \noalign{\smallskip} 
 \hline 
 \noalign{\smallskip} 
FL8Y~J0324.7$+$3411         & 1H~0323$+$342                       	       &51.19	&34.20  & 0.061  & PL    &2.00$\times10^{-11}$	&436	& 2.93  & -      &  1                       \\  
FL8Y~J0850.0$+$5108         & SBS~0846$+$513                     	       &132.51	&51.14  & 0.585  & LP    &7.83$\times10^{-12}$	&638	& 2.12  & 0.10 &  1                        \\  
    \hspace{24pt} High State  &       -                                                      	       & -	        &-	     &     -     & PL    &1.08$\times10^{-10}$	&300   	& 2.10  & -      &  2                       \\
FL8Y~J0932$+$5306    & NVSS~J093241$+$530633                           &143.17  &53.11  &  0.597 & PL    &1.10$\times10^{-11}$      &300	&2.39   & -       & 3                       \\
3FGL~J0937.7$+$5008$^{\mathrm{a,b}}$  &  GB6~J0937$+$5008            &144.30   &50.15 &  0.276 & PL    &8.00$\times10^{-12}$      &300	 &2.41  & -       &  3                      \\  
FL8Y~J0948.9$+$0022                     &  PMN~J0948$+$0022               &147.24  &0.37 & 0.585  & BKPL  &1.06$\times10^{-10}$	 &1000	 &2.30  & 3.40  &  4                       \\  
   \hspace{24pt}     High State          &     -                                             	       &  -         &  -    &      -    & PL     &9.60$\times10^{-10}$	 &300	 &2.55  & -       &  5                       \\
    \hspace{20pt}  ``Flare'' State       &    -                                              	 &   -        &  -    &      -    & PL     &2.88$\times10^{-9}$	         &300	 &2.55  & -       &  6                        \\
FL8Y~J0958.0$+$322$^{\mathrm{c}}$&  CRATES~J095821$+$ 32235  &149.52 &32.37 & 0.530 & PL     &1.76$\times10^{-12}$	         &538	&2.73  & -        &  1                       \\
J1102$+$2239                         &  -                                                   &165.70 &22.63  & 0.453&PL      &1.40$\times10^{-10}$           &300      &3.10  & -       & 7                       \\   
J1222$+$0413                         &  -                                                   &185.64 & 4.21   &0.966 &PL      &2.01$\times10^{-11}$           &444      &2.87  & -       & 1$^{\mathrm{d}}$  \\
J1246$+$0238                         &  -                                                   &191.83 &2.53    &0.363 &PL      &1.18$\times10^{-10}$           &300      &3.10  & -       & 7                       \\  
FL8Y~J1305.2$+$5108  & SDSS~J130522.74$+$511640.2$^{\mathrm{e}}$ &196.31 &51.14 &0.785 &PL  &1.57$\times10^{-12}$      &437      &2.91   & -      & 1                       \\  
FL8Y~J1331.0$+$3031                        &  3C~286 $^{\mathrm{f}}$       &202.75 &30.53   & 0.850 &PL      &9.37$\times10^{-14}$         &1445     &2.37   & -      & 1                        \\ 
NVSS~J142106$+$385522$^{\mathrm{g}}$    &    -                           & 215.28 &  38.92 &  0.489 & PL  &4.00$\times10^{-12}$           &300	 &2.66  & -      & 3                       \\  
FL8Y~J1443.1$+$4729 & B3~1441$+$476$^{\mathrm{e}}$               &220.80 & 47.49  & 0.706 & PL    &1.08$\times10^{-12}$	         &614        &2.65 & -      & 1                        \\  
FL8Y~J1505.0$+$0326 & PKS~1502$+$036	                              &	226.26 &	3.44	   & 0.408  & PL   &1.11$\times10^{-11}$	&506	 &2.67  & -      &   1                       \\  
 \hspace{24pt} High State  &     -                               	                            &	-          &      -     & -        & PL   &1.4$\times10^{-9}$	        &250 	 &2.54  & -      &   8                       \\
3FGL~J1520.3$+$4209$^{\mathrm{a}}$ & TXS~1518$+$423            &  230.17 & 42.19   & 0.484 & PL   &7.50$\times10^{-12}$         &300	 &2.67  & -      &   3                       \\
SDSS~J164100.10$+$345452.7  &         -                                      &250.25   &34.91    & 0.164 & PL   &1.2$\times10^{-11}$            &300	 &2.5   & -       &  9                       \\   
FL8Y~J1644$+$2618        &FBQS~J1644.9$+$2619                      &251.24  &26.31   & 0.145    &PL    &2.02$\times10^{-12}$          &549      & 2.74 & -       & 1                        \\ 
\hspace{24pt} High State   & -                                                                    & -         & -         & -            &PL    &5.00$\times10^{-11}$          &300      &2.50  & -      &  10                      \\
\hspace{24pt} Flare State   &  -                                                                   & -         & -         & -            &PL    &6.35$\times10^{-10}$          &300      &2.50  & -      &  10                      \\
FL8Y~J2007.9$-$4432 & PKS~2004$-$447  	                            & 301.98 &$-$44.55 &0.240 & PL  &4.72$\times10^{-12}$	         &578	 &2.65 & -       &  1                       \\  
3FGL~J2118.4$+$0013$^{\mathrm{a}}$ & PMN~J2118$+$0013       & 319.57  & 0.22   &  0.463   & PL   &3.20$\times10^{-12}$          &300	  &2.23 & -      &  3                       \\
FL8Y~J2119.2$-$0728 & AT20G~J211853$-$073227                  &319.81  &$-$7.48  & 0.260 & PL   &3.46$\times10^{-12}$	 &452	  &2.81 & -      &   1                       \\
 \noalign{\smallskip} 
  \noalign{\smallskip}  
  \hline
\end{tabular}
\end{center}
\begin{list}{\it Notes.}{} 
  \item {Redshift are drawn from NED. $\gamma$-ray spectral models: PL=power law, LP=log-parabola, BKPL=broken power law. }
  \item[$^{\mathrm{a}}$]{Previously mis-classified as FSRQs \citep{Paliya2018:nls1_sloan}. }
  \item[$^{\mathrm{b}}$]{Classified as candidate NLS1 by \citet{Paliya2018:nls1_sloan}, due to its relatively weak \ion{Fe}{II} emission (\ion{Fe}{II}/\Hbeta = 0.05). }
  \item[$^{\mathrm{c}}$]{Assumed associated with NVSS~J095820$+$322401 \citep{Paliya2018:nls1_sloan}. }
  \item[$^{\mathrm{d}}$]{Also see \citet{Yao2015:1222+0413}. }
  \item[$^{\mathrm{e}}$]{Also see \citet{Liao2015:arxiv}. }
  \item[$^{\mathrm{f}}$]{Also see \citet{Berton2017:3C286}.}
  \item[$^{\mathrm{g}}$]{Classified as candidate NLS1 by \citet{Paliya2018:nls1_sloan}, due to the incompleteness in its 
                             \Hbeta{} emission line profile, leading to the ambiguity in the FWHM measurement.}
\end{list}   
\begin{list}{\it References.}{} 
\item For the models we adopted:  
(1) {\it Fermi-}LAT 8-year Source List (FL8Y); 
(2) \citet[][]{Paliya2016:0846}; 
(3) \citet[][]{Paliya2018:nls1_sloan}; 
(4) \citet[][]{Abdo2009:J0948discov}; 
(5) \citet[][]{Foschini2011:J0948}; 
(6) this work: flaring state, assumed a factor of 3 brighter than the high state; 
(7) \citet[][]{Foschini2011:nlsg.conf}; 
(8) \citet[][]{Dammando2016:1502}; 
(9) \citet[][]{Lahteenmaki2018};
(10) \citet[][]{Dammando2015:1644+2619}. 
\end{list}   
\end{table*}

\setcounter{table}{1} 
 \begin{table*}  
\tiny
 \tabcolsep 1pt  
 \begin{center} 
 \caption{Setup of the ({\tt ctools}) simulations: 
site, IRF, exposure time, number of realisations run ($N_1$) for the detection in the full band (20--150\,GeV), 
and number $M$ of additional bins over which detection was performed ($N_2$ realisations), their exposure and energy ranges (see Sect.~\ref{romano_nls1:sims_dets}). 
\label{romano_nls1:table:sims}} 
  \begin{tabular}{lccccc | rccll } 
 \hline 
 \noalign{\smallskip} 
   Source Name                  & \hspace{-3pt}Site$^{\mathrm{a}}$ & IRF                                        & Exp.  & Sim.  & Energy\hspace{+2pt}     & Bins       & Exp.  &  Sim.   & Energy  Bands & \\ 
        (Model)                        &                             &                                              & (h)      &  $N$  & (GeV)       &  $M$     & (h)     &  $N_2$ & (GeV)   &         \\
\noalign{\smallskip} 
 \hline 
 \noalign{\smallskip} 
J0324$+$3410                      &N&	{\tt North\_z20\_average\_50h}	& 50 &	100	& 20--150  & 	&  & &	 	\\
 \noalign{\smallskip}
J0849$+$5108  High               &N &	{\tt North\_z20\_average\_50h}	& 50  &	1000 & 20--150 & 3 & 50 &1000 &20--30, 30--50, 50--150 &  \\  
 \noalign{\smallskip}
J0932$+$5306                       &N	&	{\tt North\_z20\_average\_50h}	& 50 &	100	& 20--150  & 	&  & &	 	\\ 
J0937$+$5008                    &N	&	{\tt North\_z20\_average\_50h}	& 50 &	100	& 20--150  & 	&  & & 	 	\\ 
 \noalign{\smallskip}
J0948$+$0022                          &N	&	{\tt North\_z20\_average\_5h}	& 3 &	1000 & 20--150 &3 & 3 &1000 &20--30, 30--50, 50--150  & \\  
  \hspace{10pt}  (``Flare'' State)&S	&	{\tt South\_z20\_average\_5h}	& 3 &	1000 & 20--150 &3 & 3 &1000 &20--30, 30--50, 50--150  & \\  
                                                  &N	&	{\tt North\_z20\_average\_5h}	& 5 &	1000 & 20--150 &3 & 5 &1000 &20--30, 30--50, 50--150  & \\  
                                                  &S	&	{\tt South\_z20\_average\_5h}	& 5 &	1000 & 20--150 &3 & 5 &1000 &20--30, 30--50, 50--150  & \\  
                                                  &N	&	{\tt North\_z20\_average\_5h}	& 10 &	1000 & 20--150 &3 & 10 &1000 &20--30, 30--50, 50--150 & \\  
                                                  &S	&	{\tt South\_z20\_average\_5h}	& 10 &	1000 & 20--150 &3 & 10 &1000 &20--30, 30--50, 50--150 & \\  
\noalign{\smallskip}
J0948$+$0022                           &N&	{\tt North\_z20\_average\_5h}	& 5 &	1000 & 20--150 &3  & 5  &1000  &20--30, 30--50, 50--150 &  \\  
       \hspace{14pt}  (High State) &S	&	{\tt South\_z20\_average\_5h}	& 5 &	1000 & 20--150 &3  & 5  &1000  &20--30, 30--50, 50--150  & \\  
                                                  &N	&	{\tt North\_z20\_average\_5h}	& 10 &	1000 & 20--150 &3 & 10 &1000 &20--30, 30--50, 50--150 &  \\  
                                                  &S	&	{\tt South\_z20\_average\_5h}	& 10 &	1000 & 20--150 &3 & 10 &1000 &20--30, 30--50, 50--150 & \\  
                                                  &N	&	{\tt North\_z20\_average\_50h}	& 50 &	1000 & 20--150 &4 & 50 &1000 &20--30, 30--50, 50--150, 20--50 &  \\  
                                                  &S	&	{\tt South\_z20\_average\_50h}	& 50 &	1000 & 20--150 &4 & 50 &1000 &20--30, 30--50, 50--150, 20--50  & \\ 
\noalign{\smallskip}
J0948$+$0022                         &N	&	{\tt North\_z20\_average\_50h}	& 100 &	1000 & 20--150 &1 &100&1000 &20--50 & \\  
   \hspace{16pt} (Quiescent)     &S	&	{\tt South\_z20\_average\_50h}	& 100 &	1000 & 20--150 &1 &100&1000 &20--50 & \\  
 \noalign{\smallskip}
J0958$+$3224	                &N	&	{\tt North\_z20\_average\_50h}	     & 50 &	100	& 20--150    & 	&  & &  	\\
J1102$+$2239                        &N 	&{\tt North\_z20\_average\_50h} &  50 &	100	& 20--150    & 	&  &  &    \\  
J1222$+$0413   	                &N 	&{\tt North\_z20\_average\_50h} & 50 &	100	& 20--150    & 	&  &  &   \\ 
                                                &S 	&{\tt South\_z20\_average\_50h} & 50 &	100	& 20--150    & 	&  &  &    \\  
J1246$+$0238                       &N 	&{\tt North\_z20\_average\_50h} & 50 &	100	& 20--150    & 	&  &  &    \\  
                                               &S 	&{\tt South\_z20\_average\_50h} & 50 &	100	& 20--150    & 	&  &  &     \\  
J1305$+$5116	               &N	&{\tt North\_z20\_average\_50h}& 50  &	100	& 20--150    & 	&  & &   	\\

J1331$+$3030                          &N 	&{\tt North\_z20\_average\_50h} & 50 &	100	& 20--150     & 	&  &  &   \\
J1421$+$3855                            & N	&{\tt North\_z20\_average\_50h}	& 50 &	100	& 20--150     & 	&  & &   	\\ 

J1443$+$4725	                   &N	&	{\tt North\_z20\_average\_50h}	& 50 &	100	& 20--150  & 	&  & &  	\\ 
J1505$+$0326  	                   &N	&	{\tt North\_z20\_average\_50h}	& 50 &	100	& 20--150  & 	&  & &    	\\
         \hspace{20pt} (Quiescent) &S	&	{\tt South\_z20\_average\_50h}	& 50 &	100	& 20--150  & 	&  & &    	\\
J1520$+$4209                          &N	&	{\tt North\_z20\_average\_50h}	& 50 &	100	& 20--150  & 	&  & &  	\\ 
J1641$+$3454	                   &N&	{\tt North\_z20\_average\_50h}	& 50 &	100	& 20--150  & 	&  & &  	\\  
 \noalign{\smallskip}
J1644$+$2619 Flare                     &N 	& {\tt North\_z20\_average\_5h} & 10 &	100	& 20--150     &   & \\  
\hspace{20pt} (High)                    &N & {\tt North\_z20\_average\_50h} & 50 &	100	& 20--150     &   & \\                                    
\hspace{20pt} (Quiescent)            &N &{\tt North\_z20\_average\_50h} & 50 &	100	& 20--150     & 	&  &  &    \\                    
 \noalign{\smallskip}
J2007$-$4434                            &S	&	{\tt South\_z20\_average\_50h}	& 50 &	100	& 20--150  & 	&  & & 	\\ 
J2118$+$0013                           &N	&	{\tt North\_z20\_average\_50h}	& 50 &	100	& 20--150  & 	&  & &  	\\ 
                                                  &S	&	{\tt South\_z20\_average\_50h}	& 50 &	100	& 20--150  & 	&  & &  	 	\\  
J2118$-$0732	                          &S	&	{\tt South\_z20\_average\_50h}	& 50 & 100	& 20--150  & 	&  & &   		\\ 
\noalign{\smallskip}
  \hline
  \noalign{\smallskip}
No cut-off$^{\mathrm{b}}$ \\ 
\noalign{\smallskip}
  \hline
  \noalign{\smallskip}
J0849$+$5108  (High)      &N &	{\tt North\_z20\_average\_5h}&    &	  &  & 8 & 10 &1000 &20--30, 30--50, 50--75, 
                                                                                                                                                                        75--100, 100--140, 140--200, 200--300, 300--400    &  \\
                                            &N &	{\tt North\_z20\_average\_50h}&    &	  &  & 8 & 50 &1000 &20--30, 30--50, 50--75, 
                                                                                                                                                                        75--100, 100--140, 140--200, 200--300, 300--400    &  \\
J0849$+$5108 (Quiesc.) &N	&	{\tt North\_z20\_average\_50h}	& 100 &	1000 & 20--150 &1 &100&1000  &20--50                               & \\ 
 \noalign{\smallskip}
 J0948$+$0022           &N	&	{\tt North\_z20\_average\_5h}& 3 &1000 & 20--150 &8 & 3 &1000 &20--30, 30--50, 50--75, 
                                                                                                                                                                        75--100, 100--140, 140--200, 200--300, 300--400    &   \\
  \hspace{10pt}  (``Flare'' State)   &S	&	{\tt South\_z20\_average\_5h}& 3 &1000 & 20--150 &8 & 3 &1000 &20--30, 30--50, 50--75, 
                                                                                                                                                                        75--100, 100--140, 140--200, 200--300, 300--400    &   \\ 
   \noalign{\smallskip}
                                                      &N	&	{\tt North\_z20\_average\_5h}& 5 &1000 & 20--150 &3 & 5 &1000 &20--30, 30--50, 50--150   & \\
                                                       &S	&	{\tt South\_z20\_average\_5h}& 5 &1000 & 20--150 &3 & 5 &1000 &20--30, 30--50, 50--150   & \\
  \noalign{\smallskip}
J0948$+$0022$^{\mathrm{b}}$           &N      &	{\tt North\_z20\_average\_5h}& 5 &1000 & 20--150 &8  &5  &1000  &20--30, 30--50, 50--75, 
                                                                                                                                                                        75--100, 100--140, 140--200, 200--300, 300--400    &   \\  
       \hspace{10pt}  (High State)  &S	&	{\tt South\_z20\_average\_5h}& 5 &1000 & 20--150 &8  & 5  &1000  &20--30, 30--50, 50--75, 
                                                                                                                                                                        75--100, 100--140, 140--200, 200--300, 300--400    &   \\   
  \noalign{\smallskip}

J1505$+$0326 	               &N	&  	{\tt North\_z20\_average\_5h} &   &	  &   &8  &5  &1000  &20--30, 30--50, 50--75, 
                                                                                                                                                                        75--100, 100--140, 140--200, 200--300, 300--400    &   \\  
\hspace{10pt}  (High State)     &S	&  	{\tt South\_z20\_average\_5h} &   &	  &   &8  &5  &1000  &20--30, 30--50, 50--75, 
                                                                                                                                                                        75--100, 100--140, 140--200, 200--300, 300--400    &   \\  
 \noalign{\smallskip}
  \hline
  \end{tabular}
\end{center}
\begin{list}{}{} 
   \item[$^{\mathrm{a}}$]{CTA site selected for the simulations: N=North (La Palma), S=South (Paranal).}
  \item[$^{\mathrm{b}}$]{The input model did not include the cut-off due to internal absorption (see Sect.~\ref{romano_nls1:discussion}).}
\end{list}   
  \end{table*}

%
Narrow-Line Seyfert 1 galaxies (NLS1s) are a subclass of active galactic nuclei (AGN) 
characterised in the optical regime by 
  narrow permitted emission lines (\Hbeta{} \FWHM $<$ 2000 $\kms$, \citealt[][]{Goodrich1989:nls1def}),  
  weak forbidden [\ion{O}{III}] lines ([\ion{O}{III}]\,$\lambda 5007/ \Hbeta < 3$),
  and strong Iron emission lines (high \ion{Fe}{II}/\Hbeta,  \citealt[][]{OsterbrockP1985:nls1def}). 
As such, these galaxies are located at the lower end of the line-width distribution for the Seyfert 1 population, 
thus distinguished from the bulk of Seyfert 1 galaxies (broad-line Seyfert 1s, BLS1s). 
In the X-rays NLS1s have equally extreme properties, as they show rapid and large-amplitude variability 
\citep[][]{BollerBF1996:softX}, with some showing X-ray flares up to a factor of 100 in flux, on timescales of days,  
compared to the factors of  a few seen in BLS1s. 
These distinctive properties can be understood in terms of lower masses for the central black hole
($10^6$--$10^8$ M$_{\sun}$) 
compared to BLS1s with similar luminosities and higher accretion rates, close to the Eddington limit
\citep[e.g.][]{Peterson2004}. 

Traditionally, NLS1s are considered hosted in spiral/barred galaxies 
\citep[][]{Crenshaw2003:hosts}, and generally not strong radio emitters, 
but evidence has been collected that a small fraction 
(4--7\,\%, \citealt[][]{Komossa2006:rlnl1q,Cracco2016}) 
of NLS1s are radio loud and show a flat radio spectrum 
(\citealt[][]{Oshlack2001:pks2004-447,Zhou2003:0948,Yuan2008}; 
see also, \citealt[][]{Lahteenmaki2017}).
Furthermore, a hard component was found in the {\it Swift}/XRT X-ray spectra 
of NLS1s, as well as spectral variability in the hard X-ray as observed by INTEGRAL/IBIS and {\it Swift}/BAT 
\citep[][]{Foschini2009:Adv}. 
These properties are strongly reminiscent of those of jetted sources \citep[see, e.g.][]{Foschini2012:review,Foschini2015:fsrl_nls1,Dammando2016:jets_nls1}.

The first detection by {\it Fermi}-LAT  of a NLS1 in the $\gamma$-rays (E $> 100$\,MeV), 
PMN~J0948$+$0022 \citep[][]{Abdo2009:J0948discov,Foschini2010:J0948}, 
and subsequent follow-ups \citep[][]{Abdo2009:J0949mw,Foschini2011:J0948} 
confirmed that its multi-wavelength behaviour was that of a 
source with a relativistic jet, like those observed in blazars. 
Since then, a total of 20 sources identified as NLS1s 
have been found by {\it Fermi}-LAT to emit in the  $\gamma$-rays  
and the sample is bound to grow in time. 
However, currently no firm detection has been obtained in the very high energy (VHE) regime. 
Indeed, \citet[][]{Falcone2004} found marginal evidence for flaring (at the 2.5$\sigma$ level) 
but did not detect significant emission from 1H~0323$+$342 with Whipple above 400\,GeV.
Also, VERITAS observations of PMN~J0948$+$0022 (5\,hr) only yielded upper limits 
at E $> 100$\,GeV \citep[][]{Dammando2015:J0948}.  
A third NLS1, PKS 2004$-$447 was observed but not detected at VHE by H.E.S.S.
\citep[][]{2014A&A...564A...9H}. 
The detection in the VHE regime would provide important clues on the location of the emitting region, 
since the central region of  NLS1s, analogously to FSRQ, are expected to be highly opaque to 
gamma rays above few tens of GeV.

The future of NLS1s science in the VHE regime 
will benefit from the construction of the Cherenkov Telescope Array (CTA) 
\citep[][]{2011ExA....32..193A,2013APh....43....3A}, 
which will afford us a wide (20\,GeV--300\,TeV)  energy range. 
The CTA array will include different classes of telescopes, i.e., 
the large-sized telescopes (LSTs, diameter D$\sim23$\,m),   
the medium-sized telescopes (MSTs, D$\sim12$\,m)  
and the small-sized telescopes (SSTs, primary mirror D$\sim4$\,m).   
The full array will be installed in two sites, one for each hemisphere to allow an 
all-sky coverage. 
The baseline CTA setup ~\citep[][]{2017AIPC.1792b0014H,2017Msngr.168...21H} is composed of  
a Northern site, 
located at  the Observatorio del Roque de los Muchachos on the island of La Palma (Spain) 
where 4 LSTs and 15 MSTs, covering an area of $\sim1$\,km${^2}$, will be installed, 
and a Southern site, located at the European Southern Observatory's (ESO's) Paranal Observatory 
in the Atacama Desert (Chile), that will  cover an area of about 4\,km$^{2}$, where 
4 LSTs, 25 
 MSTs, and 70 SSTs will be installed. 
CTA will provide an average differential sensitivity a factor 5--20 better with respect 
to the current imaging atmospheric Cherenkov telescope (IACT) arrays; in particular
for transients and flaring events (time-scales of $\sim 1$ day or shorter) 
CTA will be about two orders of magnitude more sensitive with
respect to {\it Fermi}-LAT at the overlapping energy of 25\,GeV, 
thus allowing an unprecedented opportunity to investigate flaring $\gamma$-NLSy1 galaxies. 

In this paper we consider all currently known $\gamma$-ray emitting NLS1s and 
explore the prospects for observations of the whole sample with CTA. 
In Sect.~\ref{romano_nls1:sample} we define our sample of NLS1s, in 
Sect.~\ref{romano_nls1:simulations} we describe our simulation setup, in 
Sect.~\ref{romano_nls1:results} we present our results and in Sect.~\ref{romano_nls1:discussion} 
discuss their implications.

 	 \section{Data Sample}  \label{romano_nls1:sample}
%
 Our sample (Table~\ref{romano_nls1:table:sample}) 
consists of all objects classified as NLS1s that have been detected in the 
gamma-rays, as mainly reported by the {\it Fermi-}LAT 8-year Source List 
(FL8Y, {\tt gll\_psc\_8year\_v3.fit} v.\ 
2018-01-03)\footnote{\href{https://fermi.gsfc.nasa.gov/ssc/data/access/lat/fl8y/}{https://fermi.gsfc.nasa.gov/ssc/data/access/lat/fl8y/}.}
and in the existing literature. 
Although the sample is not complete in the statistical sense, 
since it is not characterised by a flux limit, it does include all $\gamma$-NLS1s 
(as well as two candidates, see Notes in Table~\ref{romano_nls1:table:sample}) 
identified at the time of writing. 

Table~\ref{romano_nls1:table:sample}  
includes, for each source, coordinates (Equatorial, J2000, Cols.~3, 4) and redshift (Col.~5) as provided by 
the NASA/IPAC Extragalactic Database (NED)\footnote{\href{https://ned.ipac.caltech.edu}{https://ned.ipac.caltech.edu}.}.
It also reports the spectral parameters for the best fit models to the {\it Fermi} data that we 
adopted for each source (and flux state, Cols.~6--10), and the reference from which it was drawn or derived (Col.~11).
The spectral models are, \\ 
{\it i)} a power law (PL),  
\begin{equation}
\frac{dN}{dE}=K_0 \left( \frac{E}{E_0} \right) ^{-\Gamma},
\end{equation}
where $K_0$ is the normalisation (in units of ph\,cm$^{-2}$\,s$^{-1}$\,MeV$^{-1}$)
$E_0$ is the pivot energy (in MeV), 
and  $\Gamma$ is the power-law photon index;  \\
{\it ii)} a log-parabola (LP) 
\begin{equation}
\frac{dN}{dE}=K_0 \left( \frac{E}{E_0} \right) ^{-\alpha -\beta \ln \left(E/E_0\right)},
\end{equation}
where $K_0$ is the normalisation, $E_0$ is the pivot energy,  
$\alpha$ is the spectral slope, $\beta$ the curvature;  \\
{\it iii)} a broken power-law (BKPL) 
\begin{equation}
\frac{dN}{dE} = K_0 \times 
               \begin{cases}
                 \left(\frac{E}{E_b}\right)^{-\Gamma_1}      & \quad {\rm if \,\,\,} E < E_b \\
                 \left (\frac{E}{E_b}\right)^{-\Gamma_2}      & \quad {\rm otherwise }, 
               \end{cases}
\end{equation}
where $K_0$ is the normalisation, and  $\Gamma_1$ and $\Gamma_2$ 
are the spectral indices at energies lower and higher than the break energy $E_b$.

\begin{figure*} 
\vspace{-2.5truecm}
\centerline{
    \hspace{-0.5truecm} 
                \includegraphics*[angle=0,width=18.8cm]{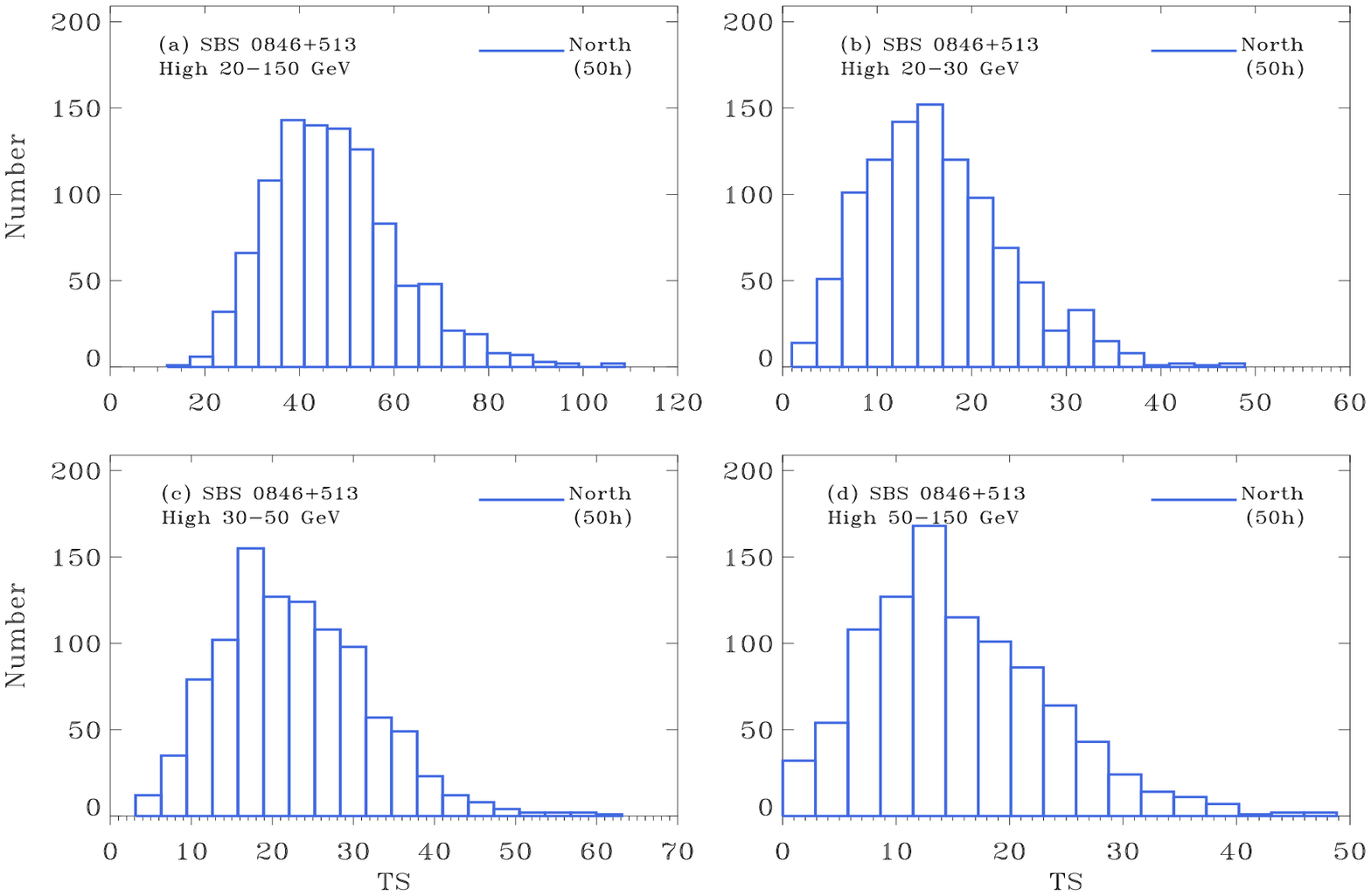}
    \vspace{-12.5truecm} 
}
\caption{Distribution of the test statistic (TS) values for \sbs{} in the high state in 50\,hr. 
See Table~\ref{romano_nls1:table:dets} for details. }
\label{romano_nls1:fig:dets_over_T_0850_high} 
\end{figure*} 

 	 \section{Simulations}  \label{romano_nls1:simulations}
%
%
The simulations were performed with the  {\tt ctools}  \citep[][v.\ 
1.4.2]{Gammalib_ctools_2016}\footnote{\href{http://cta.irap.omp.eu/ctools/}{http://cta.irap.omp.eu/ctools/}. }  
 analysis package  and the public CTA instrument response 
files\footnote{\href{https://www.cta-observatory.org/science/cta-performance/}{https://www.cta-observatory.org/science/cta-performance/}.  }
(IRF, v.\ prod3b-v1). 
Each source is assumed to be observed from the site that provides the largest source elevation, 
computed from the difference between the geographic latitudes of the CTA sites 
(North latitude: 28.76 N; South latitude: 24.68 S) 
and the source declination (Table~\ref{romano_nls1:table:sample}, Col. 4); 
accordingly, the corresponding prod3b-v1 IRFs (reported in Table~\ref{romano_nls1:table:sims}, Col.~3) 
were used for the simulations.

In the model definition XML file for  {\tt ctools}, the spectral model component was defined as a 
{\tt FileFunction} type, so that the spectrum was provided as an ASCII file containing 
energy (in MeV) and differential flux values (in units of ph\,cm$^{-2}$\,s$^{-1}$\,MeV$^{-1}$), 
described according to 
\begin{equation}
M_{\rm spectral}\left(E\right)   =  N_0 \frac{dN}{dE}, 
\end{equation}
where $N_0$ is the normalisation. 

The input spectral models have been derived by extrapolating the best-fit {\it Fermi} spectra 
(the parameters are reported in Table~\ref{romano_nls1:table:sample}) to the CTA energy range, 
including the effects of the gamma-ray absorption both 
along the path to the Earth (which, at the relevant energies, 
is due to the interaction with the UV-optical part of the extragalactic background light, EBL), 
and inside the source (internal absorption).
The correction for absorption by EBL (providing substantial attenuation only above about 100\,GeV) 
has been applied to all spectra by using the model of \citet[][]{Dominguez2011:EBL}. 

Absorption of gamma rays within the source itself is expected because of the interaction with the 
UV ambient radiation (originating in the accretion disk and in the broad line region, 
see e.g. \citealt{PoutanenStern2010}). 
Because of the presence of the prominent $Ly_{\alpha}$ line of Hydrogen, the most relevant spectral 
feature induced by internal absorption is a marked drop at $\approx 20$--30\,GeV. 
Due to the lack of a detailed physical and geometrical modelling of each source (and each state), 
in particular because of the currently unconstrained location of the gamma-ray emitting region, 
for this paper we chose to mimic the drop with a simple analytical description, a 
cut-off at 30\,GeV ($\propto e^{-E/E_{\rm cut}}$, $E_{\rm cut}=30$\,GeV), 
while in  future planned works we shall investigate the effects of more realistic BLR absorption models. 
There is indeed evidence of photons being detected at energies 
in excess of 10\,GeV by {\it Fermi} from some of our sources, e.g.\ 
J0324$+$3410 \citep[up to 32.7\,GeV,][]{Paliya2015:Fermi_rl_nls1},
\sbs{} \citep[16.5\,GeV,][]{Sahakyan2018},   
\pks{} \citep[21.1\,GeV,][]{Dammando2016:1502}. 
We applied such cut-off to all sources characterised by an unbroken power law in the LAT band. 
The cut-off was not considered for the cases in which the LAT spectrum is reproduced by a log-parabola, 
already characterised by an intrinsic curvature leading to the progressive softening of the spectrum 
(see Table~\ref{romano_nls1:table:sample} and notes on individual objects 
below)\footnote{We note that we shall also consider a more optimistic scenario, i.e., no internal absorption, for  \sbs. 
}.

We considered only the instrumental background included in the IRFs 
({\tt CTAIrfBackground}) and no further contaminating astrophysical sources in the 
5\,deg field of view (FOV) we adopted for event extraction. 

By default, energy dispersion is not considered in the {\tt ctools} fits, but 
because of the spectral softness of NLS1s, our investigation of their detectability   
was also carried out at energies well below 100\,GeV, where the effects of the 
energy dispersion can become important \citep[][]{Maier2017:cta_performance}. 
Inclusion of the energy migration matrix in our simulations ({\tt edisp=yes})  
especially when performing likelihood analysis, involves 
computation times up to 10 times longer for the ranges of spectral parameters and exposures we considered. 
Therefore, after performing several test runs, we decided not to include the effects of 
energy dispersion in this exploratory work.
As we show in Appendix~\ref{romano_nls1:appendix_edisp}, 
given the exposure times selected and the resulting detection significance 
of our sources, we are confident that the effects are not significant enough to 
change our conclusions. 

As a test case for relatively faint sources, we generally selected an exposure of 50\,hr,
but considered exposures as short as 3\,hr for flaring states, 
and as long as 100\,hr for quiescent states 
(details in Table~\ref{romano_nls1:table:sims}, Col.~4). 
We note that 50\,hr correspond on average to the expected exposure that CTA can accumulate in 
one observing year on a single source, while 3--5\,hr correspond to the integration of 1--2\,days, 
depending on source visibility and target scheduling.

In the following, we discuss details of the inputs for specific sources 
for which more than one flux state was considered.

\begin{figure} 
\vspace{-0.4truecm}
\centerline{
    \hspace{+0.3truecm} 
                 \includegraphics[angle=0,width=9.3cm]{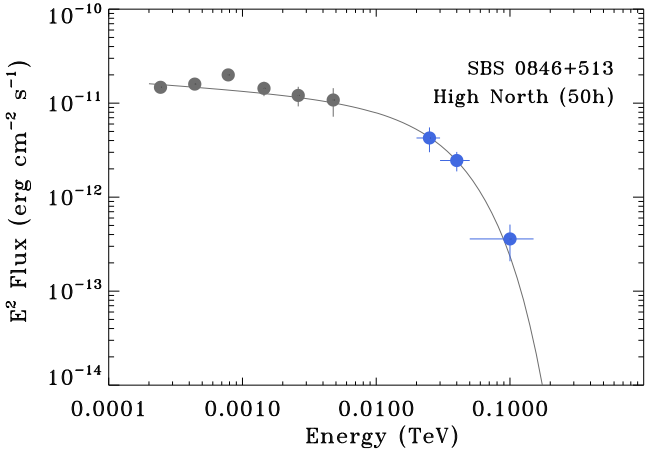}
}
\caption{SED of \sbs{} in the high state. The grey line is the input model, 
the blue points the simulated fluxes for 50\,hr of exposure. 
The grey points are from \citet[][F2 flare]{Paliya2016:0846} . 
}
\label{romano_nls1:fig:sed_0850_high} 
\end{figure} 

\begin{figure*} 
\vspace{-2.5truecm}
\centerline{
 \hspace{-0.5truecm} 
                 \includegraphics*[angle=0,width=18.8cm]{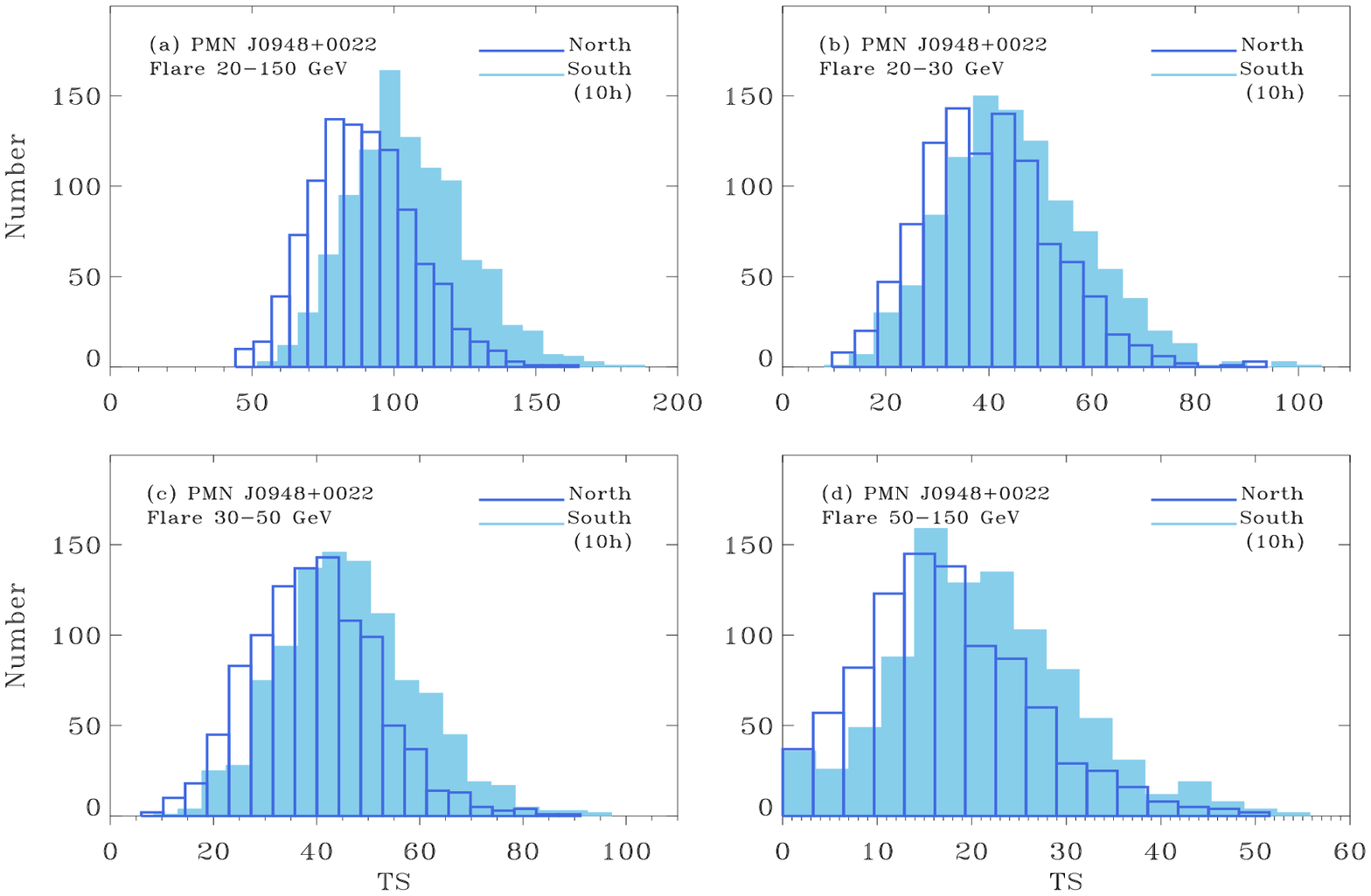}
\vspace{-12.5truecm} 
}
\caption{Distribution of the TS values for \pmn{} in the ``flare'' state in 10\,hr.  
See Table~\ref{romano_nls1:table:dets} for details. 
The systematic shift in the TS distribution to larger mean TS values for the South is related to the 
slightly larger sensitivity of the South array with respect to the North array 
(see \href{https://www.cta-observatory.org/science/cta-performance}{https://www.cta-observatory.org/science/cta-performance}).
}
\label{romano_nls1:fig:dets_over_T_0948_flare} 
\end{figure*} 

 {\it J0849$+$5108 (\sbs). }
Two flux states were considered for this source. 
The high-state one \citep[F2 flare,][integrated over 120\,days]{Paliya2016:0846} is modelled by means of a simple power-law model with photon 
index 2.10 and an integrated gamma-ray flux ($0.1<E<300$\,GeV) of $9.92\times 10^{-8}$\,ph\,cm$^{-2}$\,s$^{-1}$.
The average flux state has been drawn from the FL8Y list, assuming a log-parabola spectrum 
(see FL8Y on-line FITS file for the spectral parameter) and an integrated gamma-ray flux ($1<E<100$\,GeV) 
of $2.18\times 10^{-9}$\,ph\,cm$^{-2}$\,s$^{-1}$.
The high state model was corrected for EBL absorption and intrinsic (BLR) absorption (cut-off at 30\,GeV), 
the average state model was only corrected for EBL. 

 {\it J0948$+$0022 (\pmn). }  
Three flux states were considered for this source. 
The quiescent state ($F_{\rm E>200 MeV} =(3.9\pm0.3)\times10^{-8}$\,ph\,cm$^{-2}$\,s$^{-1}$) 
was derived from \citet[][integrating over 5 months]{Abdo2009:J0948discov}, and is described 
by a broken power law with photon indices $\Gamma_1=2.3$ and $\Gamma_2=3.4$ and a break at 1\,GeV. 
The high state ($F_{\rm E>100 MeV}=(1.02\pm0.02)\times10^{-6}$\,ph\,cm$^{-2}$\,s$^{-1}$)  
is described by a simple power-law model with photon index 
$\Gamma=2.55$ \citep[][]{Foschini2011:J0948}.  
A third, flaring state was defined as three times brighter than the high state,
with the same spectral shape. 
All models were corrected for EBL absorption  
and intrinsic (BLR) absorption (cut-off at 30\,GeV). 

 {\it FL8Y~J1505.0$+$0326 (\pks).}  
We considered two flux states for this source, the quiescent state being derived from FL8Y. 
From \citet[][]{Dammando2016:1502}, instead,   
we drew a high state (as observed on 2015 December 20, 1\,day integration) 
described by a power-law with a photon index $\Gamma = (2.54 \pm 0.04)$ and
a flux $F_{\rm 0.1<E<300 GeV} = (93 \pm 19) \times 10^{-8}$\,ph\,cm$^{-2}$\,s$^{-1}$. 
For this particular flare we assume, as also concluded by \citet[][]{Dammando2016:1502},
based on the observed 3-week delay between the $\gamma$ and radio light curve (15\,GHz) peaks, 
that the dissipation region may lie outside the BLR. 
Therefore no cut-off was applied to the input model for our simulations of the high state, 
while the  average state model was corrected for EBL absorption and intrinsic (BLR) absorption (cut-off at 30\,GeV). 

{\it J1644$+$2619 (FBQS J1644.9$+$2619).} 
Three flux states were considered for this source, the 
quiescent state being derived from FL8Y. 
From \citet[][]{Dammando2015:1644+2619}, instead 
we drew a high state as an average over 2012 July 15  to October 12,  
described by a power-law with a photon index $\Gamma = (2.5 \pm 0.2)$ and
a flux $F_{\rm 0.1<E<100 GeV} = (5.2 \pm 1.0) \times 10^{-8}$\,ph\,cm$^{-2}$\,s$^{-1}$ 
and a flaring  state as a daily average obtained on 2012 August 18 (MJD 56157)
with a flux $F_{\rm 0.1<E<100 GeV} = (66 \pm 22) \times 10^{-8}$\,ph\,cm$^{-2}$\,s$^{-1}$. 
For sake of simplicity, we assumed for this flaring state the same photon index
reported for the high state.
 All models were corrected for EBL absorption  
and intrinsic (BLR) absorption (cut-off at 30\,GeV).

 	 \subsection{Detectability}   \label{romano_nls1:sims_dets}
%
A first set of simulations was dedicated to ascertain whether the sources would be 
detectable by CTA. The general setup is summarised in Table~\ref{romano_nls1:table:sims}. 
In the following, we shall consider the reliability of a source detection in an energy band 
based on the test statistic \citep[TS,][]{Cash1979,Mattox1996:Egret} of the 
maximum likelihood model fitting. 
In particular, the detection will have a high significance when $TS \geq 25$  \citep[][]{Mattox1996:Egret} 
and a low significance when $10 \leq TS < 25$. 
The source will not be considered detected for $TS <10$ and an upper limit will need to be calculated instead. 

Given the spectral softness of NLS1s, to investigate their detectability   
we selected a soft energy band, that is, 20--150\,GeV, 
in which the LSTs provide the full system sensitivity. 
In this band we used the task {\tt ctobssim} to create event lists based on our 
input models, including the randomised background events. 
We then used the task {\tt ctlike} to fit a power-law model 
$M_{\rm spectral}(E)=k_0 \left( \frac{E}{E_0} \right) ^{\Gamma}$, 
where $k_0$ is the normalisation (or {\tt Prefactor}, in units of ph\,cm$^{-2}$\,s$^{-1}$\,MeV$^{-1}$)
$E_0$ is the pivot energy ({\tt PivotEnergy} in MeV), 
and  $\Gamma$ is the power-law photon index ({\tt Index}). 
In the fits we left {\tt Prefactor} and {\tt Index} free to vary while we kept 
{\tt PivotEnergy}  fixed at 100\,GeV.    
The task {\tt ctlike} uses maximum likelihood model fitting and 
calculates TS.

\begin{figure*} 
\vspace{-2.5truecm}
\centerline{
 \hspace{-0.5truecm} 
                 \includegraphics*[angle=0,width=18.8cm]{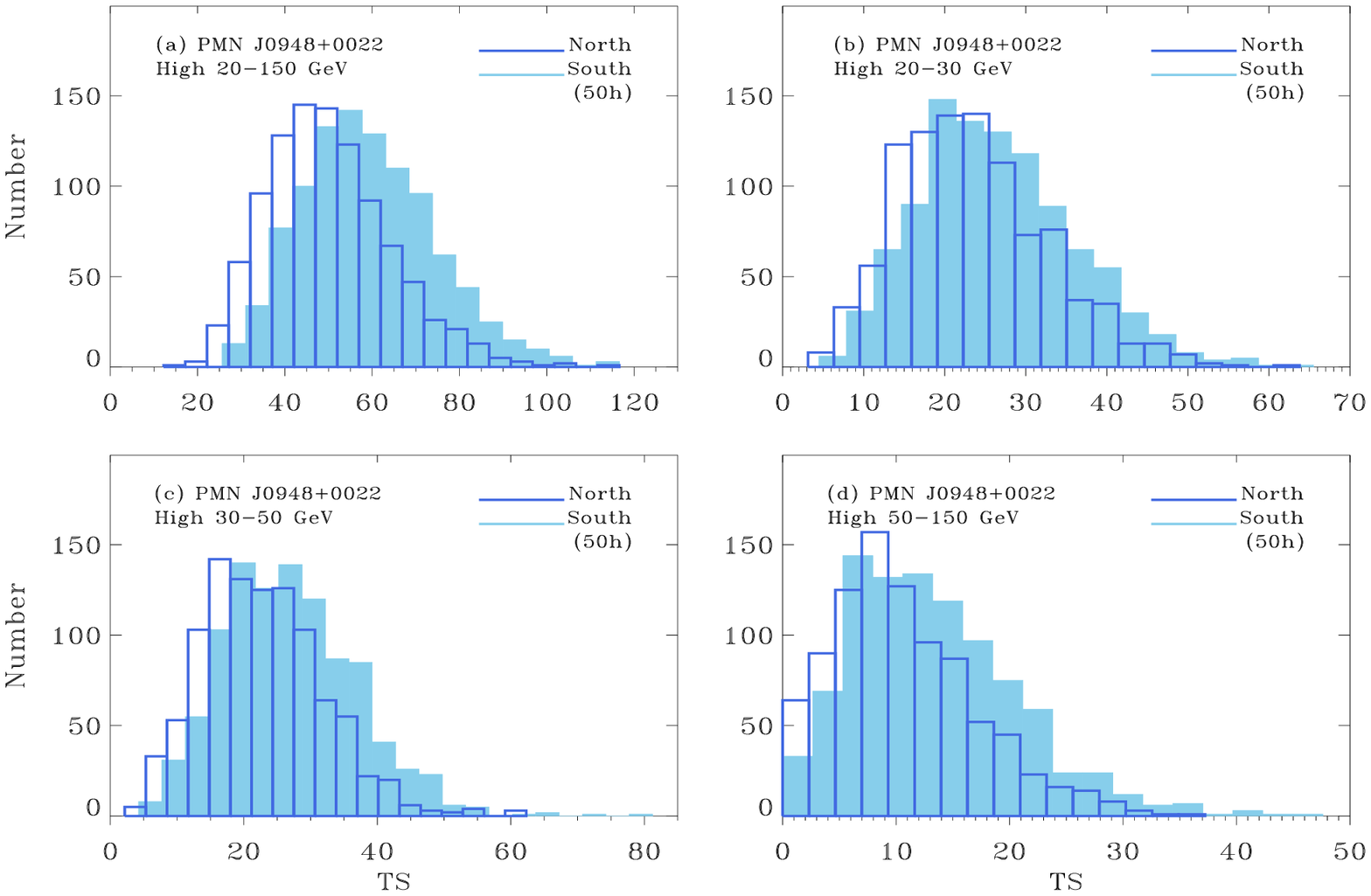}
\vspace{-12.5truecm} 
}
\caption{Distribution of the TS values for \pmn{} in the high state in 50\,hr.  
See Table~\ref{romano_nls1:table:dets} for details. } 
\label{romano_nls1:fig:dets_over_T_0948_high} 
\end{figure*} 

\begin{figure*} 
\vspace{-8.5truecm}
\centerline{
    \hspace{-0.5truecm} 
                 \includegraphics*[angle=0,width=18.8cm]{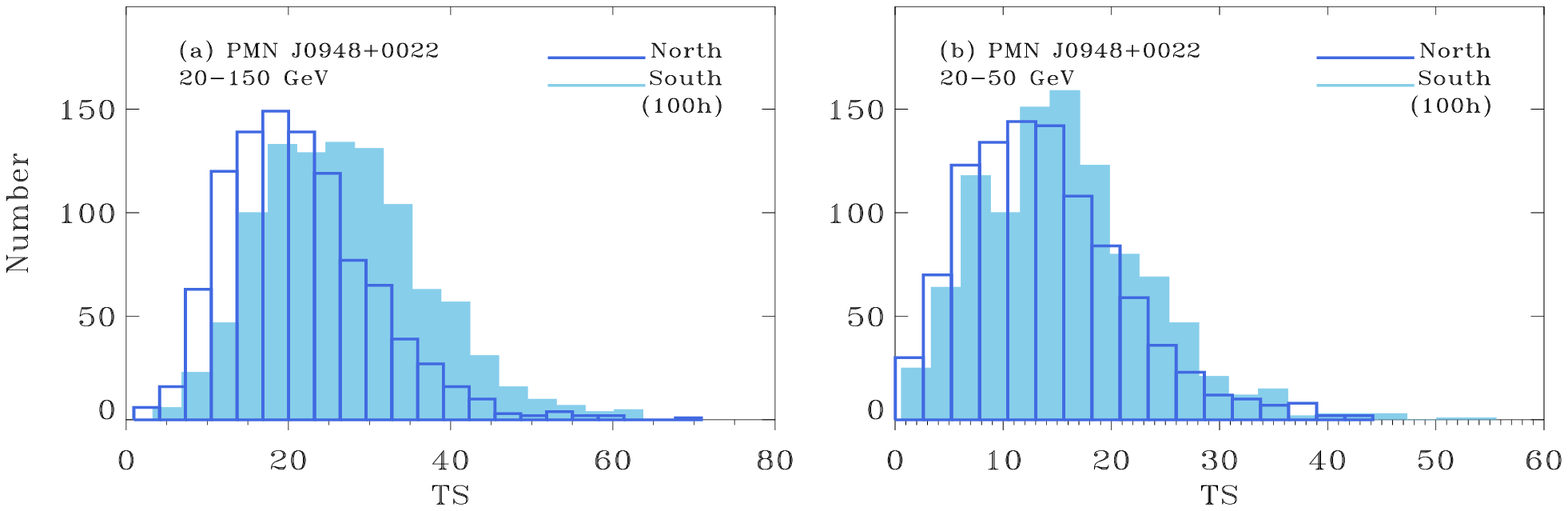}
    \vspace{-12truecm} 
}
\caption{Distribution of the TS values for \pmn{} in the quiescent state in 100\,hr.  
See Table~\ref{romano_nls1:table:dets} for details.}
\label{romano_nls1:fig:dets_over_T_0948_q} 
\end{figure*} 

\begin{figure} 
\vspace{-0.4truecm}
\centerline{
 \hspace{-0.1truecm}
                \includegraphics[angle=0,width=9.cm]{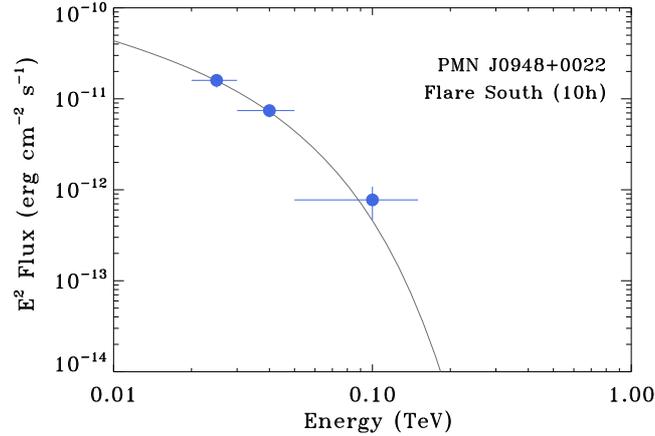}
}
\caption{SED of \pmn{} in flare. The grey line is the input model, 
the blue points the simulated fluxes (10\,hr exposure). 
}
\label{romano_nls1:fig:sed_0948_flare} 
\end{figure} 

\begin{figure} 
\vspace{-0.4truecm}
\centerline{
                 \includegraphics[angle=0,width=9.3cm]{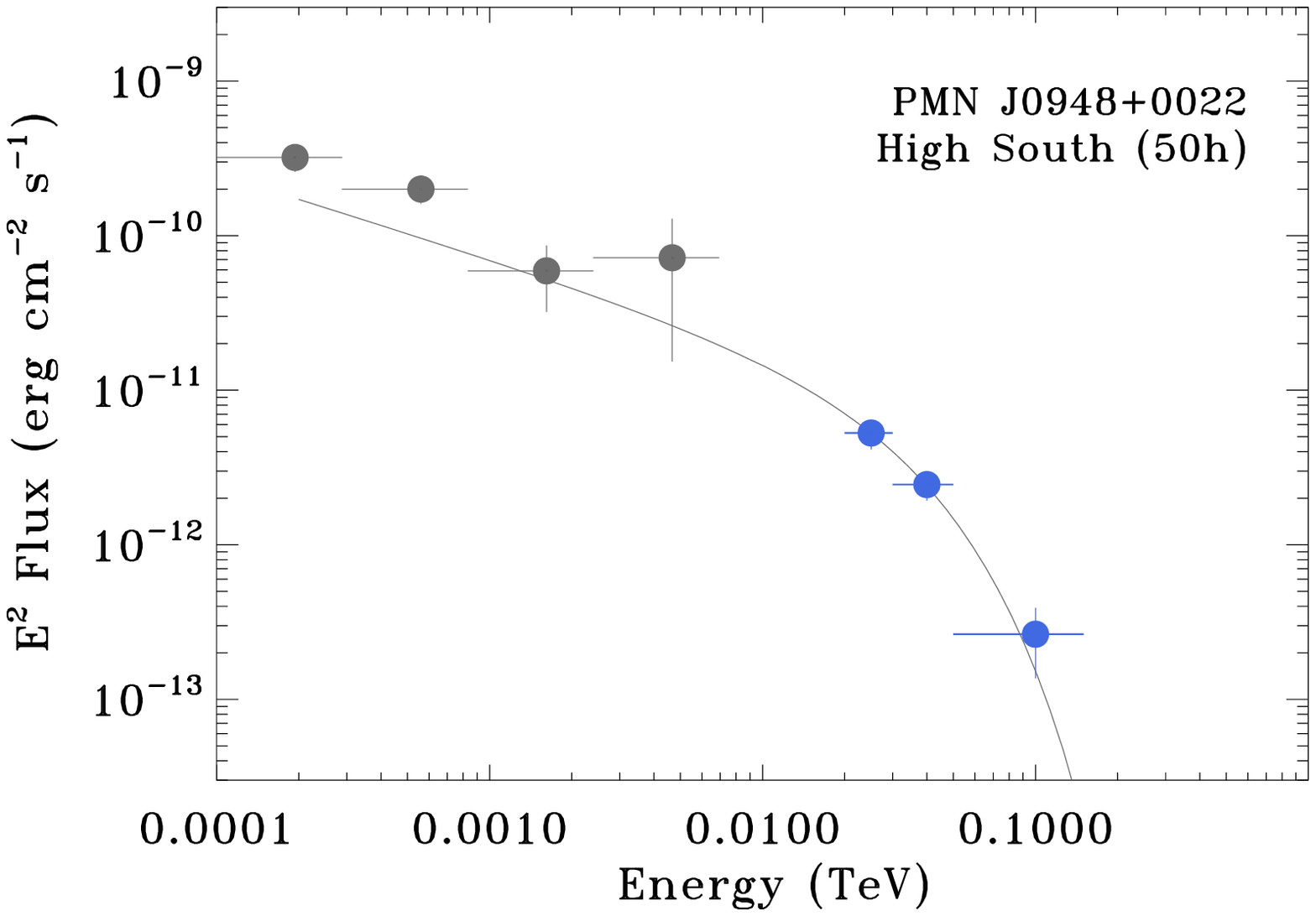}
}
\caption{SED of \pmn{} in the high state. 
The grey line is the input model, 
the blue points the simulated fluxes (50\,hr exposure). 
The grey points are from \citet[][]{Foschini2011:J0948} . 
}
\label{romano_nls1:fig:sed_0948_high} 
\end{figure} 

To reduce the impact of variations between individual realisations 
\citep[see, e.g.][]{Gammalib_ctools_2016} 
we performed sets of $N$ (Table~\ref{romano_nls1:table:sims}, Col.~5) 
statistically independent realisations by adopting different seeds  ({\tt seed}) 
for the randomisation, where $N$ was chosen as a compromise between 
accuracy in the assessment of the detection confidence level and computing 
time\footnote{In order 
to efficiently run such large number of simulations,
we performed them through Amazon Web Services, 
following the methods described in  Landoni et al. (2018, in prep). }. 
We thus obtained a set of $N$ values of TS. 
We then derived the percentage of the detections for $TS >10$ (Table~\ref{romano_nls1:table:dets}, Col.~5)
and the percentage of the detections for $TS >25$ (Table~\ref{romano_nls1:table:dets}, Col.~6). 
These represent the detection confidence levels.  
Then, the mean TS value and its uncertainty were calculated as the mean, 
$\overline{TS_{\rm sim}}  = \frac{1}{N}\sum_{k=1}^{N}TS_{\rm sim}(k)$, 
and square root of the standard deviation of the sample of $N$ values, 
$s^2_{\rm sim}=\frac{1}{N-1}\sum_{k=1}^{N}(TS_{\rm sim}(k)-\overline{TS_{\rm sim}})^2$.
They are reported in Table~\ref{romano_nls1:table:dets} (Col.~7). 
For each realisation the best fit spectral parameters were used to calculate 
$N$ values of flux in the 20--150\,GeV energy band. 
Similarly, the flux mean and uncertainty were calculated and are reported in 
Table~\ref{romano_nls1:table:dets} (Col.~8).  
%
When the source was not detected, 
we calculated the 95\,\% confidence level upper limits on fluxes 
by using the task {\tt ctulimit} (see Table~\ref{romano_nls1:table:uls}).  
As inputs we used the first event file generated with {\tt ctobssim} 
for which the task {\tt ctlike} converged ($TS>0$),  
and a model obtained by fitting the absorbed data with log-parabola model.

 	 \subsection{Spectral properties}  \label{romano_nls1:sims_spectra}
%
For the sources that were detected (Table~\ref{romano_nls1:table:dets}), 
we then proceeded to investigate their spectral properties. 
We considered a set of $M$ energy bins (Table~\ref{romano_nls1:table:sims}, Col.~7) 
covering an energy band reported in Table~\ref{romano_nls1:table:sims} (Col.~10), namely, 
soft (20--30\,GeV), mid (30--50\,GeV), softmid (20--50\,GeV), and hard (50--150\,GeV). 
In each bin we used the task {\tt ctobssim} to create event lists, then 
used the task {\tt ctlike} to fit each spectral bin with a power-law model
with the same set-up as for the detections (Sect.~\ref{romano_nls1:sims_dets}), 
with {\tt PivotEnergy} fixed at 
25\,GeV for the soft band, 
45\,GeV for the mid band,
35\,GeV in the softmid band, 
and 100\,GeV for the hard band. 

For each source we obtained sets of $N_2$ realisations 
(Table~\ref{romano_nls1:table:sims}, Col.~8). We then proceeded 
as in Sect.~\ref{romano_nls1:sims_dets} 
and calculated average TS and spectral parameters and 95\,\% confidence level upper limits 
(see Table~\ref{romano_nls1:table:uls}).

\begin{figure} 
\vspace{-0.4truecm}
\centerline{
 \hspace{+1.2truecm}
                 \includegraphics[angle=90,width=9.0cm]{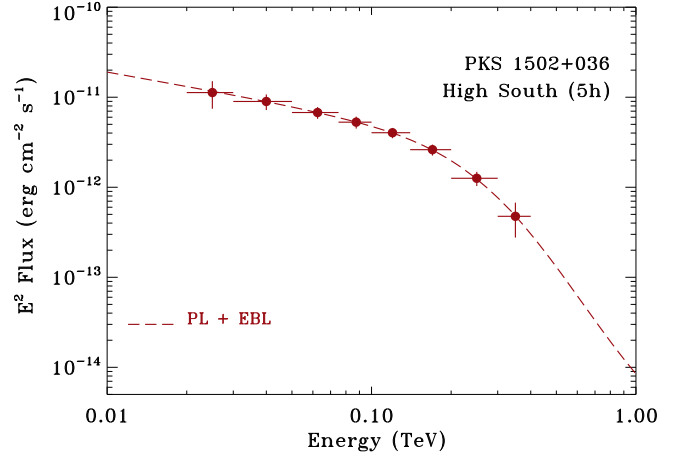}
}
\caption{SED of \pks{} in the high state for an exposure of 5\,hr. 
The red dashed line is the input model which does not include the cut-off due to internal absorption, 
the red  points the simulated fluxes (Table~\ref{romano_nls1:table:dets2}). 
}
\label{romano_nls1:fig:sed_1505_high_nocut5} 
\end{figure} 

 	 \section{Results}  \label{romano_nls1:results}

 	 \subsection{\sbs{}} \label{romano_nls1:res_0850}

Fig.~\ref{romano_nls1:fig:dets_over_T_0850_high}a shows the distributions of the TS  for 
\sbs{} in the high state, in the full energy band (20--150\,GeV), while  
Fig.~\ref{romano_nls1:fig:dets_over_T_0850_high}(b,c,d) shows the distributions of the TS  for 
high state in the narrower energy bands. 
Table~\ref{romano_nls1:table:dets} reports the percentage of the detections for $TS >10$ and  for $TS >25$ (Col.~5, 6), 
and the mean TS value (Col.~7) based on our simulations, as well as the mean flux in each of the energy bands we considered (Col.~8). 
We find that this source is 
\begin{itemize}
\item detected in the high state \citep[as described by][]{Paliya2016:0846} in 50\,hr 
          (Fig.~\ref{romano_nls1:fig:sed_0850_high}); 
\item not detected in quiescence (FL8Y) in 100\,hr (even though no cut-off at 30\,GeV representing internal absorption was applied, 
see Table~\ref{romano_nls1:table:dets2}). 
\end{itemize}

 	 \subsection{\pmn{}}  \label{romano_nls1:res_0949}

In Fig.~\ref{romano_nls1:fig:dets_over_T_0948_flare}, 
           \ref{romano_nls1:fig:dets_over_T_0948_high}, and
           \ref{romano_nls1:fig:dets_over_T_0948_q} (Panels a)
we plot the distributions of the TS  for \pmn{} in the full energy band (20--150\,GeV) 
while in the flaring, high and quiescent states, respectively.  
In Fig.~\ref{romano_nls1:fig:dets_over_T_0948_flare},
            \ref{romano_nls1:fig:dets_over_T_0948_high} (Panels b,c,d), and 
            \ref{romano_nls1:fig:dets_over_T_0948_q} (b) 
we plot the distributions of the TS for \pmn{}
in the narrower bands in the flaring, high and quiescent states. 
The percentages of the detections for $TS >10$ and  for $TS >25$, mean TS 
and mean flux in each of the energy bands we considered can be found in 
Table~\ref{romano_nls1:table:dets}. 

This source, therefore, is 
\begin{itemize}
\item detected in the ``flare'' state in all bands in 10\,hr  
               (Fig.~\ref{romano_nls1:fig:sed_0948_flare})
\item detected in the ``flare'' state up to 50\,GeV in 3\,hr;  
\item detected in high state \citep[as described by][]{Foschini2011:J0948} in all bands in  50\,hr
               (Fig.~\ref{romano_nls1:fig:sed_0948_high}); 
\item detected in high state up to 50\,GeV in 10\,hr; 
\item detected in quiescence \citep[as described by][]{Abdo2009:J0948discov}  in the total band and 
          softmid (20--50\,GeV) band in 100\,hr. 
\end{itemize}

 	 \subsection{\pks{}} \label{romano_nls1:res_1502}
Since \pks{} was particularly bright during the high state, partly due to the fact that no cut-off at
30\,GeV was applied, for the high state we performed a test for detection in 8 bands, extending up to 400\,GeV
(see Table~\ref{romano_nls1:table:sims}). 
We find that this source is 
\begin{itemize}
\item detected in the high state \citep[as described by][]{Dammando2016:1502} in 5\,hr in all bands up to 400\,GeV
          (Fig.~\ref{romano_nls1:fig:sed_1505_high_nocut5}, Table~\ref{romano_nls1:table:dets2}); 
          we note, again, that no cut-off at 30\,GeV was applied in this case (see Sect.~\ref{romano_nls1:simulations}); 
\item not detected in quiescence (FL8Y) in 100\,hr (Table~\ref{romano_nls1:table:uls}). 
\end{itemize}

 	 \subsection{Other sources} \label{romano_nls1:res_other}
 
We investigated the possibility to detect all other sources in our sample in the 20--150\,GeV energy band;
however, no  detections were obtained. 
The detailed results  
can be found in Table~\ref{romano_nls1:table:uls}, which reports 
the percentage of the detections for $TS >10$ and  for $TS >25$ (Col.~5, 6), 
and the mean TS value (Col.~7) based on our simulations, 
as well as the mean flux in each of the energy bands we considered (Col.~8). 
Col.~9, finally, reports the 95\,\% upper limits on detection.

\begin{figure} 
\vspace{-0.4truecm}
\centerline{
                 \includegraphics[angle=90,width=8.8cm]{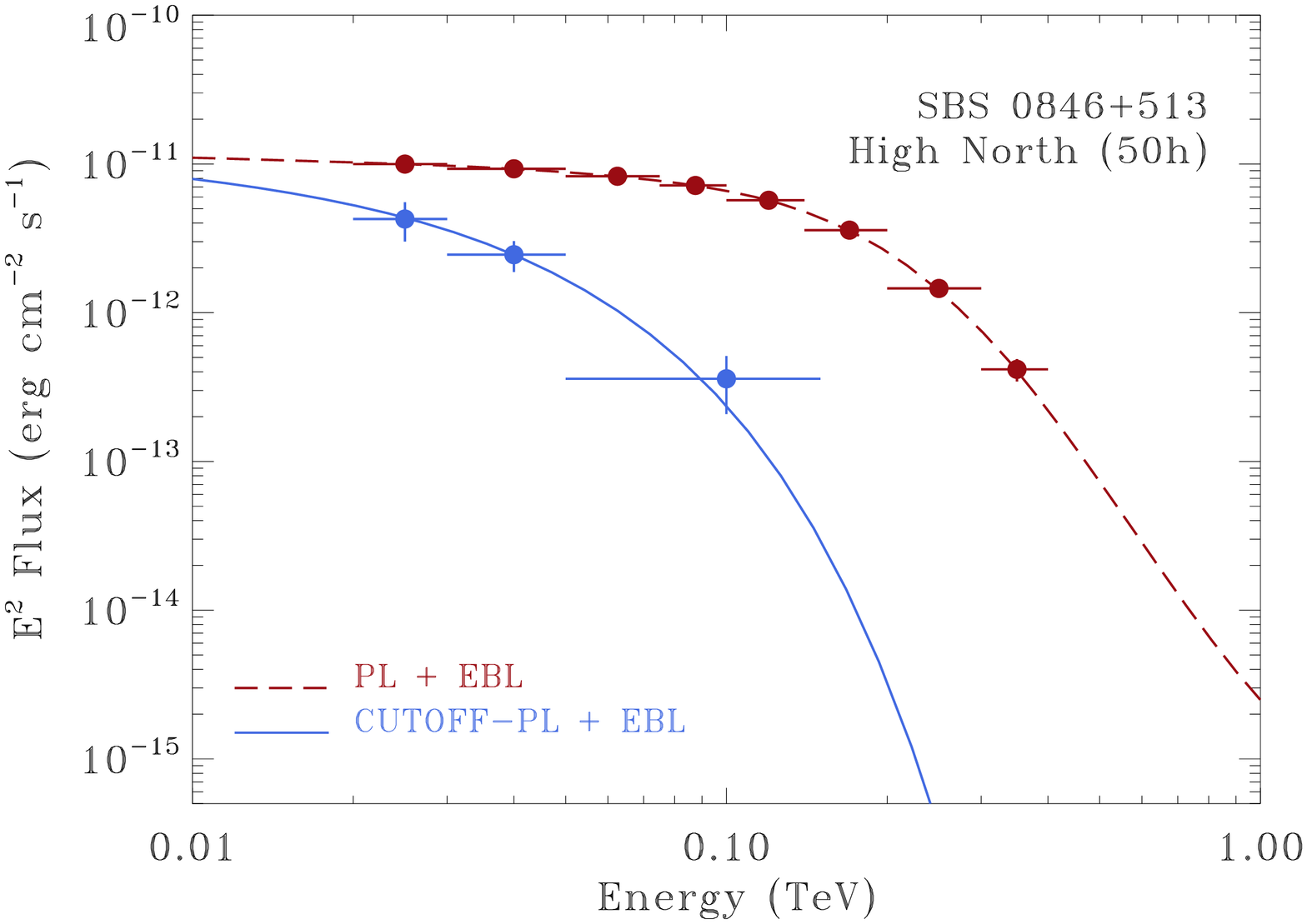}
}
\caption{SED of \sbs{} in the high state (exposure of 50\,hr). 
The blue line is the input model (see Sect.~\ref{romano_nls1:sample}), 
the blue points the simulated fluxes (Table~\ref{romano_nls1:table:dets}). 
The red dashed line is the input model which does not include the cut-off due to internal absorption, 
the red  points the simulated fluxes (Table~\ref{romano_nls1:table:dets2}). 
}
\label{romano_nls1:fig:sed_0850_high_nocut50} 
\end{figure} 
\begin{figure} 
\vspace{-0.4truecm}
\centerline{
                  \includegraphics[angle=90,width=8.8cm]{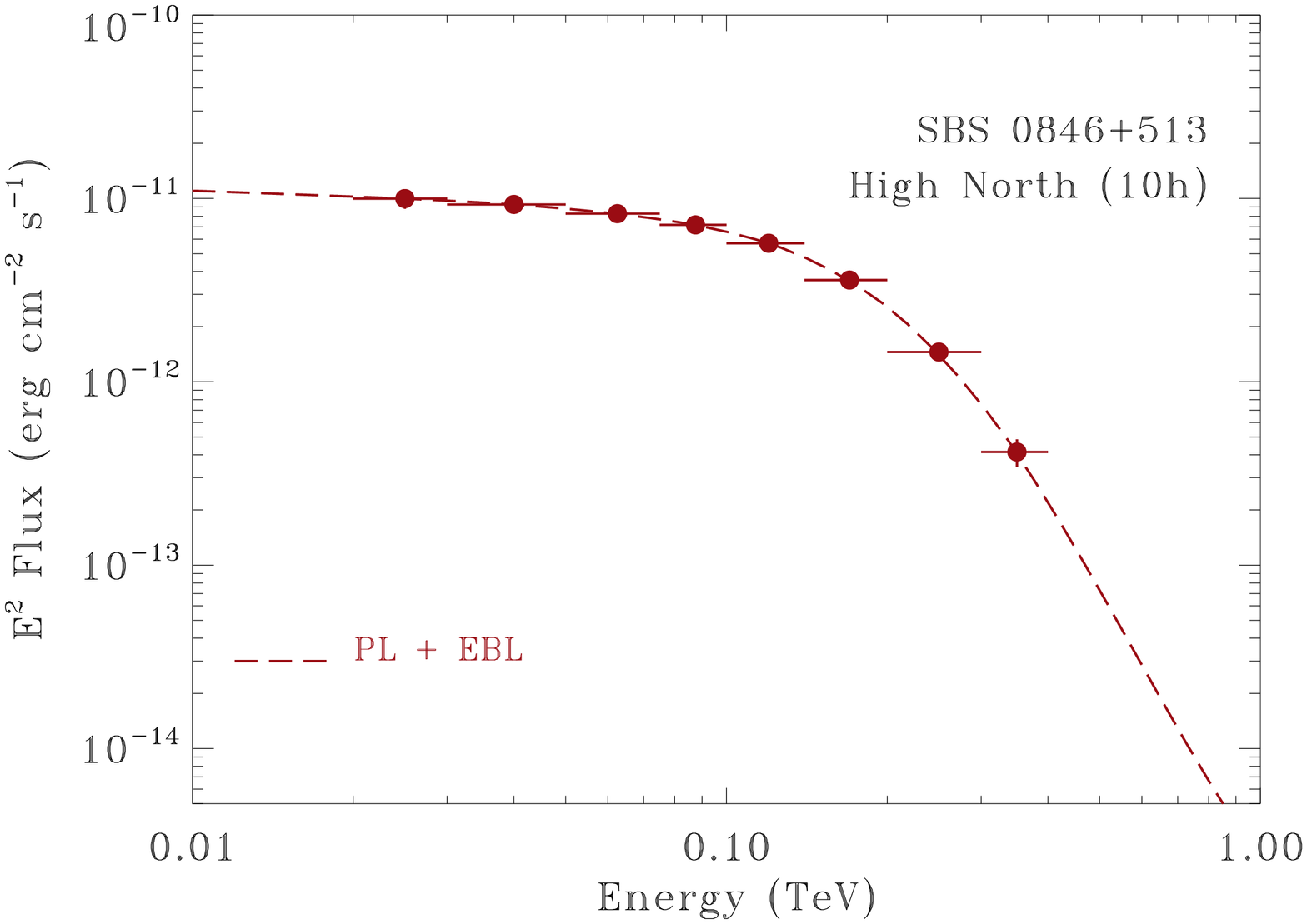}
}
\caption{SED of \sbs{} in the high state for an exposure of 10\,hr. 
The red dashed line is the input model which does not include the cut-off due to internal absorption, 
the red  points the simulated fluxes (Table~\ref{romano_nls1:table:dets2}). 
}
\label{romano_nls1:fig:sed_0850_high_nocut10} 
\end{figure} 

 	 \section{Discussion}  \label{romano_nls1:discussion}

In this paper we performed an investigation of the largest sample of $\gamma$-NLS1s to date, some in several flux states, 
in order to assess their suitability as potential CTA targets 
and to provide guidance in the possible observing strategy. 

A first set of simulations was dedicated to ascertain whether the sources would be 
detectable by CTA in the ``standard'' framework in which one assumed that emission 
occurs at distances from the BH smaller than the BLR radius \citep[e.g.][]{Abdo2009:J0948discov}. 
For each of the 20 sources 
we simulated event files with {\tt ctools} and performed a test for detection via the maximum likelihood method 
in the  20--150\,GeV band, the most promising one due to the relative softness of these sources. 
The main assumptions for the input spectra were that they would need to be corrected for 
absorption by EBL \citep[modelled according to ][]{Dominguez2011:EBL}, 
and intrinsic absorption  which, for simplicity, has been modelled assuming an exponential cut-off 
at 30\,GeV (see Sect.~\ref{romano_nls1:simulations}). 

As expected, due to the faintness of $\gamma$-NLS1s, we did not detect most of the sample. 
However, three sources stood out as very promising, \sbs{}, \pmn, and \pks. 
For these sources we therefore investigated their spectral properties by performing 
a detection in several energy bands. 
\sbs{} was detected in the high state, 
in 50\,hr, while  \pmn{} was  detected in the high state 
up to 150\,GeV in 50\,hr and up to 50\,GeV in 10\,hr. 
It was detected up to 150\,GeV even in quiescence 
in 100\,hr. 
\pks{} was detected in all bands up to 400\,GeV while in high state 
for which, we note, no cut-off was applied to the input model \citep[][]{Dammando2016:1502}.  
This exploratory work, therefore, demonstrates that  $\gamma$-NLS1s are indeed promising 
CTA targets even when the input spectra are heavily absorbed by EBL and intrinsic absorption. 
Furthermore, we note that the number of sources in our sample is still small, 
and their  $\gamma$-ray duty cycle not well known. This, combined with the large uncertainties in the 
input models (in particular the location of the dissipation region in each flare, see below, and in 
\citealt[][and references therein]{Dammando2015:J0948}), may increase the 
fraction of NLS1s detected in the CTA bands.

\begin{figure} 
\vspace{-0.4truecm}
\centerline{
\hspace{+1.2truecm}
                   \includegraphics[angle=90,width=8.8cm]{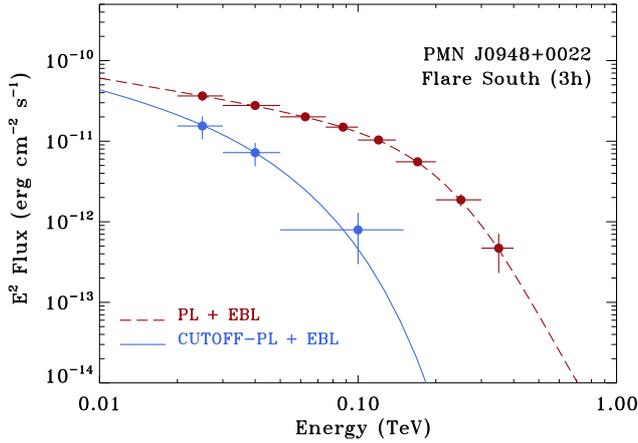}
}
\caption{SED of \pmn{} in flare (exposure of 3\,hr). 
The blue line is the input model (see Sect.~\ref{romano_nls1:sample}), 
the blue points the simulated fluxes (Table~\ref{romano_nls1:table:dets}). 
The red dashed line is the input model which does not include the cut-off due to internal absorption, 
the red  points the simulated fluxes (Table~\ref{romano_nls1:table:dets2}). 
}
\label{romano_nls1:fig:sed_0948_flare_nocut3} 
\end{figure} 

Evidence is emerging that for blazars the location of the gamma-ray emitting region 
may not always be placed at the same distance from the central black-hole during 
different flaring episodes of the same source 
as suggested by,  e.g., 
\citet[][]{Foschini2011:FermiSymp} for PKS 1222+216
(and subsequently by, e.g., \citet[][for PKS~1510-089]{Brown2013:pks1510}, 
\citet[][]{Coogan2016:3C454} and \citet{Finke2016:3C454} for 3C~454.3).  
This is especially supported by the absence in some FSRQs of the expected spectral breaks/cut-off 
\citep[][and references therein]{VERITAS2015:pks1441,Costamante2018:gammas}
at 20--30\,GeV expected to mark the absorption of the gamma rays with the UV radiation emitted by the BLR clouds 
\citep[e.g.][]{PoutanenStern2010}.
It is also supported by the detection of seven FSRQs in the VHE band 
(\citealt[][]{Sitarek2015:0218+35,Cerruti2017:arXiv170800658C,Mirzoyan:2017ton0599,Mukherjee2017:2017ton0599,Neronov2010:pks1221+216,MAGIC2011:pks1221+21,Albert2008:3C279,Ahnen2015:pks1441,VERITAS2015:pks1441,HESS2013:pks1510-089}; 
also see TeVCat\footnote{\href{http://tevcat.uchicago.edu/}{http://tevcat.uchicago.edu/}. } 
for further references). 
Support to this also comes from the dramatic change of the position of the synchrotron and inverse Compton 
peaks for some FSRQ during extreme flares 
\citep[][]{Ghisellini2013:pmn2345,Pacciani2014:blazar_zone,Ahnen2015:pks1441}, 
interpreted as due to the smaller cooling suffered by the electrons in the less dense radiation field outside the BLR. 
In fact, the lower cooling would allow the acceleration mechanism to push the electrons at larger energies, determining 
the shift of the spectral peaks to larger frequencies.

Due to the close similarity between blazars and NLS1s, it is conceivable that the phenomenology discussed above 
can also be displayed by NLS1s. 
We therefore investigated the impact of the position of the emitting region on the detectability for the prototypical sources 
\sbs{} and \pmn{}  by simulating a further model, in addition to those described in Sect.~\ref{romano_nls1:simulations} 
(the latter included both attenuation due to the EBL 
and an internal absorption exponential cut-off, with the exception of the high state of \pks), 
assuming that the spectrum can extend unbroken above 20--30\,GeV.  
The simulation setup is reported at the bottom of  Table~\ref{romano_nls1:table:sims}, 
the results in Table~\ref{romano_nls1:table:dets2}.

\begin{figure} 
\vspace{-0.4truecm}
\centerline{
                 \includegraphics[angle=90,width=8.8cm]{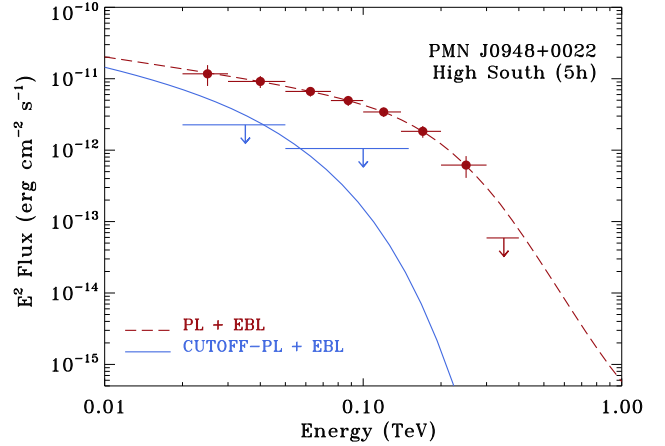}
}
\caption{Same as Fig.~\ref{romano_nls1:fig:sed_0948_flare_nocut3}  for the SED of \pmn{} in the high state (exposure of 5\,hr). 
}
\label{romano_nls1:fig:sed_0948_high_nocut5} 
\end{figure}

Figure~\ref{romano_nls1:fig:sed_0850_high_nocut50} shows the comparison 
of these two models for the high state of \sbs{} in 50\,hr, with 
the blue solid line representing the cut-off $+$ EBL model and simulated fluxes (from Table~\ref{romano_nls1:table:dets}),
while the red dashed line representing the input model which does not include the cut-off due to internal absorption
and the simulated fluxes (Table~\ref{romano_nls1:table:dets2}). 
Given the high TS obtained for each band in the latter model, 
we also simulated a 10\,hr exposure (see Fig.~\ref{romano_nls1:fig:sed_0850_high_nocut10}).
Similarly was done for \pmn, for both the ``flare'' state in 3\,hr (Fig.~\ref{romano_nls1:fig:sed_0948_flare_nocut3})
and the high state in 5\,hr (Fig.~\ref{romano_nls1:fig:sed_0948_high_nocut5}).

Figures~\ref{romano_nls1:fig:sed_0948_flare_nocut3} and \ref{romano_nls1:fig:sed_0948_high_nocut5} clearly 
show that for high gamma-ray emission states in 5 hours of observations, 
CTA will be able to discriminate between the two competing models, providing strong constraints on the location
of the jet dissipation region. For more intense gamma-ray activity (flaring state) 3 hours of observation
should allow us to perform time-selected spectroscopy of the gamma-ray event. 
We note, however, that in a more realistic situation we can expect that the spectrum, assumed here to be a power 
law with the same slope up to 1\,TeV, will display a progressive softening with energy 
(as expected, for instance, because of the transition of the IC scattering from the Thomson to the Klein-Nishina regime). 
This would have an important impact on the observed spectra, in particular at the highest energies.

$\gamma$-NLS1s are known to be quite variable on timescales of hours to days, 
timescales in which CTA has a distinct advantage over {\it Fermi}-LAT in the 20--200\,GeV 
band\footnote{\href{https://www.cta-observatory.org/science/cta-performance/\#1525680063092-06388df6-d2af}{www.cta-observatory.org/science/cta-performance/}  \#1525680063092-06388df6-d2af.}.  
$\gamma$-NLS1s, therefore, turn out to be excellent targets for observations in response to triggers from other facilities. 
As detailed in \citet[][]{ScienceCTA2017,Bulgarelli2015:ICRC,Fioretti2015:ICRC}, 
as a requirement, CTA will be able to repoint an external trigger in less than 50\,s. 
In such cases, CTA will be able to detect and obtain detailed spectra 
in a few hours for flaring states, and in a day or so for high states 
(see Fig.~\ref{romano_nls1:fig:sed_1505_high_nocut5} and \ref{romano_nls1:fig:sed_0850_high_nocut10}).

In the unfortunate circumstances of an interruption of the scientific activity of the 
current wide field of view $\gamma$-ray satellites (AGILE and {\it Fermi}-LAT) 
in combination with the possible absence of the e-ASTROGAM mission~\citep[][]{DeAngelis2017:eastrogam} 
during the CTA science phase, studying NLS1 galaxies with CTA clearly becomes paramount. 
In particular, the optimal combination of LSTs and MSTs will allow us to investigate such sources 
from a few tens up to a few hundred GeV, providing discriminating information on the location of the gamma-ray emitting region.

        \section*{Acknowledgements}

We thank R.L.C.\ Starling and F.\ D'Ammando for helpful discussions,
and  M.\ B{\"o}ttcher, M.\ Tornikoski, and J.\ Biteau as internal CTA reviewers. \\ 
%
The authors acknowledge contribution from the grant INAF CTA--SKA, 
``Probing particle acceleration and $\gamma$-ray propagation with CTA and its precursors'' (PI F.\ Tavecchio). \\ 
%
This research has made use of the NASA/IPAC Extragalactic Database (NED) which is operated 
by the Jet Propulsion Laboratory, California Institute of Technology, under contract with the 
National Aeronautics and Space Administration. \\ 
%
This research made use of {\tt ctools}, a community-developed analysis package for Imaging Air Cherenkov Telescope data. 
ctools is based on {\tt GammaLib}, a community-developed toolbox for the high-level analysis of astronomical gamma-ray data. \\ 
%
This research has made use of the CTA instrument response functions provided by the CTA Consortium and Observatory, 
see \href{https://www.cta-observatory.org/science/cta-performance/}{https://www.cta-observatory.org/science/cta-performance/} 
(version prod3b-v1) for more details. \\ 
%
We gratefully acknowledge financial support from the agencies and organizations 
listed here: 
\href{http://www.cta-observatory.org/consortium_acknowledgments}{http://www.cta-observatory.org/consortium\_acknowledgments}.
%
This paper went through internal review by the CTA Consortium.\\ 
%
We also thank the anonymous referee for swift comments that helped improve the paper.


%


\setcounter{table}{2} 
 \begin{table*} 
\small
 \tabcolsep 4pt  
 \begin{center} 
 \caption{
Results of the first set of simulations (detections in the 20--150\,GeV energy band), 
and the second set of simulations (detections in several bands). 
TS values and detection percentages and energy fluxes in each band. 
  \label{romano_nls1:table:dets}} 
  \begin{tabular}{lcccrr ccc c l } 
 \hline 
 \noalign{\smallskip} 
   Source Name                                 & CTA & Expo. &  Energy &  Det.\ c.l.$^{\mathrm{a}}$    & Det.\ c.l.$^{\mathrm{a}}$ & $\overline{TS_{\rm sim}}$  &E$^2$Flux$^{\mathrm{b}}$     & Notes            \\   
                                                         & Site &           & Range   &  (TS>10)                            & (TS>25)                         & & $\times10^{-13}$   & \\ 
                                                         &       & (hr)     & (GeV)    & (\%)                                   &   (\%)                              &            &(erg\,cm$^{-2}$\,s$^{-1}$)   &    \\
 \noalign{\smallskip} 
 \hline 
 \noalign{\smallskip} 
J0849$+$5108                                 &N 	   & 50 &   20--150 &  100.0 &   97.4 &  $47.3\pm13.8 $  &   $10.6\pm1.8 $  & Fig.~\ref{romano_nls1:fig:dets_over_T_0850_high}, 
                                                                                                                                                                                               Fig.~\ref{romano_nls1:fig:sed_0850_high}  \\
         \hspace{18pt}High State         &N 	   & 50 &   20--30 &   78.7 &   13.1 &  $16.4\pm7.6 $  &   $42.6\pm12.6 $  \\
                                            	        &N 	   & 50 &   30--50 &   94.5 &   37.7 &  $22.7\pm9.0 $  &   $24.5\pm5.7 $  \\ 
                                            	        &N 	   & 50 &   50--150 &   74.6 &   12.4 &  $15.5\pm8.1 $  &   $3.6\pm1.5 $  \\
\noalign{\smallskip} 
\noalign{\smallskip} 
J0948$+$0022                                 &N  & 3  &   20--150 &   95.8 &   59.4 &  $  28.4\pm 11.4 $  &   $    29.6\pm    8.9 $  \\ 
	\hspace{18pt}``Flare'' State	&S   & 3 &   20--150 &   97.0 &   73.5 &  $  32.9\pm 12.3 $  &   $    29.2\pm    7.9 $  \\
                                                        &N  & 3 &   20--30 &   63.8 &    6.1 &  $  13.3\pm  7.1 $  &   $   154.4\pm   52.1 $  \\
 							&S   & 3 &   20--30 &   71.5 &    9.7 &  $  14.9\pm  7.5 $  &   $   154.4\pm   48.8 $  \\
                                                        &N  & 3 &   30--50 &   63.5 &    6.7 &  $  13.3\pm  7.2 $  &   $    71.8\pm   24.3 $  \\
							&S   & 3  &   30--50 &   73.0 &   11.3 &  $  15.3\pm  7.9 $  &   $    72.1\pm   23.2 $  \\
                                                        &N  & 3  &   50--150 &  23.6 &    0.6 &  $   6.9\pm  5.1 $  &   $<30.7$ \\ 
 							&S   & 3 &   50--150 &   30.8 &    1.0 &  $   7.8\pm  5.2 $ &   $<16.8$  \\  
   \noalign{\smallskip} 
                                                        &N  & 5  &   20--150 &   99.6 &   95.7 &  $  45.8\pm 13.7 $  &   $    29.6\pm    6.1 $  \\
	                                                &S   & 5 &   20--150 &   99.7 &   98.2 &  $  53.4\pm 14.9 $  &   $    29.1\pm    5.5 $  \\
                                                        &N  & 5  &   20--30 &   90.7 &   30.3 &  $  21.0\pm  9.3 $  &   $   157.2\pm   40.5 $  \\
 							&S   & 5 &   20--30 &   94.6 &   39.6 &  $  23.6\pm  9.8 $  &   $   157.1\pm   37.3 $  \\
                                                        &N  & 5  &   30--50 &   90.0 &   28.6 &  $  20.8\pm  9.0 $  &   $    73.0\pm   19.2 $  \\
 							&S   & 5&   30--50 &   94.8 &   43.7 &  $  24.4\pm 10.0 $  &   $    73.8\pm   17.6 $  \\
                                                        &N  & 5 &   50--150 &   39.7 &    2.1 &  $   9.5\pm  6.2 $  &   $<20.3$ \\  
 							&S   & 5&   50--150 &   56.0 &    4.4 &  $  11.6\pm  6.8 $  &   $7.8\pm4.2$  \\                  
 \noalign{\smallskip} 
                                                        &N & 10 &   20--150  &100.0 &  100.0 &  $89.2\pm18.2 $  &   $29.4\pm4.2$     & Fig.~\ref{romano_nls1:fig:dets_over_T_0948_flare}, 
                                                                                                                                                                                               Fig.~\ref{romano_nls1:fig:sed_0948_flare}   \\  
	 	                                        &S  & 10 &   20--150  &100.0 &  100.0 &  $105.1\pm20.5$ &   $29.2\pm3.7$       \\
							&N & 10  &20--30        &99.9 &      89.1 &  $39.9\pm12.7$  &    $159.4\pm27.9$   \\
 							&S  & 10 & 20--30       &99.9 &      94.5 &  $45.2\pm13.8$  &    $159.6\pm26.2$   \\
                                                        &N & 10  &30--50       &99.8 &     88.8 &  $39.7\pm12.2$  &    $74.1\pm12.6$      \\
							&S  & 10 & 30--50        &99.9 &     95.5 &  $46.4\pm13.4$  &    $74.3\pm11.8$      \\
                                                        &N & 10  &50--150     &79.3 &     18.1 &  $17.3\pm8.9$    &    $8.0\pm3.6$          \\
							&S  & 10 & 50--150      &89.0 &     30.9 &  $21.1\pm9.7$    &    $7.7\pm3.1$          \\
  \noalign{\smallskip} 

 J0948$+$0022                                &N  & 5 &   20--150 &  23.1 &    0.7 &  $   7.0\pm  5.1 $ &   $<50.8 $ \\  %
	\hspace{18pt}High State	        &S   & 5 &   20--150 &   29.9 &    0.6 &  $   7.8\pm  5.2 $  &   $<33.2 $ \\  %
                                                        &N  & 5 &   20--50 &    18.6 &    0.3 &  $   6.4\pm  4.6 $  &   $ <37.7 $  \\  %
							&S   & 5  &   20--50 &    25.0 &    0.5 &  $   7.2\pm  5.1 $  &   $<22.6 $ \\ %
                                                        &N  & 5  &   50--150 &    3.7 &    0.0 &  $   3.4\pm  3.1 $  &   $ <13.1 $ \\ %
 							&S   & 5  &   50--150 &    4.7 &    0.0 &  $   3.6\pm  3.1 $  &   $ <10.5 $ \\ %
\noalign{\smallskip} 
                                                        &N  & 10  &   20--150 &   54.8 &    4.0 &  $  11.8\pm  6.8 $  &   $     9.7\pm    4.5 $  \\
							&S   & 10  &   20--150 &   66.1 &    7.3 &  $  13.7\pm  7.3 $  &   $     9.8\pm    3.9 $  \\
                                                        &N  & 10 &   20--50 &    46.5 &    2.4 &  $  10.5\pm  6.2 $  &   $    32.8\pm   13.6 $  \\ 
							&S   & 10  &   20--50 &   55.6 &    4.1 &  $  11.8\pm  6.6 $  &   $    33.1\pm   13.2 $  \\ 
                                                     &N  & 10  &   50--150 &   7.7 &    0.0 &  $   4.2\pm  3.5 $ &   $<10.3 $ \\ 
							&S   & 10 &   50--150 &   8.1 &    0.0 &  $   4.4\pm  3.6 $  &   $<7.8 $ \\ 
  \noalign{\smallskip} 
             				               &N  & 50 &   20--150  & 100.0 &   98.5 &  $49.9\pm 14.2$ &   $9.7\pm 1.8$ & Fig.~\ref{romano_nls1:fig:dets_over_T_0948_high}, 
                                                                                                                                                                                               Fig.~\ref{romano_nls1:fig:sed_0948_high}   \\ 
	 					        &S  & 50 &   20--150  & 100.0 &  100.0&  $59.3\pm 15.3$ &   $9.7\pm1.6$ & \\ 
							&N & 50   &    20--30  &   94.8 &   38.7 &  $23.5\pm9.1$    &   $52.8\pm12.1$  \\  
 							&S  & 50  &     20--30  &   97.6 &   51.5 &  $26.4\pm9.7$    &   $52.9\pm11.4$ \\  
                                                        &N & 50  &    30--50   &  94.2 &   37.8 &  $22.8\pm  9.0$   &   $24.4\pm5.7$  \\   
							&S  & 50  &    30--50   &  97.3 &   54.2 &  $26.9\pm 10.0$  &   $24.5\pm5.2$ \\   
                                                        &N & 50  &    50--150 &  47.6 &    3.2 &  $10.7\pm6.4$      &   $2.7\pm1.4$  \\   
							&S  & 50  &    50--150 &  63.5 &    7.7 &  $13.5\pm7.4$      &   $2.6\pm1.3$  \\   
 \noalign{\smallskip} 
J0948$+$0022                                &N	   & 100 & 20--150   & 93.3 &   29.1 &  $21.4\pm9.1$  &   $5.2\pm1.3$  & Fig.~\ref{romano_nls1:fig:dets_over_T_0948_q} \\  
	\hspace{18pt}Quiescent          &S 	   & 100 & 20--150   & 97.3 &   54.6 &  $26.9\pm10.2$ &  $5.3\pm1.2$  \\  
							&N      & 100 & 20--50    & 66.2 &     7.7 &  $13.8\pm7.4$  &   $12.2\pm4.6$ \\  
 							&S  	   & 100 & 20--50    & 75.7 &   11.5 &  $15.8\pm7.8$  &   $12.5\pm4.1$ \\  
  \noalign{\smallskip}
  \hline
  \end{tabular}
\end{center}
\begin{list}{\it Notes.}{} 
  \item[$^{\mathrm{a}}$]{We consider a detection to have a high significance when $TS \geq 25$  
and a low significance when $10 \leq TS < 25$. 
The source will not be considered detected for $TS <10$. }
  \item[$^{\mathrm{b}}$]{Upper limits are calculated for 95\,\% confidence level for all cases where $TS <10$.  }
 \end{list}   
  \end{table*}                
\setcounter{table}{3} 
 \begin{table*} 
\small
 \tabcolsep 4pt  
 \begin{center} 
 \caption{
Results of the first set of simulations (20--150\,GeV energy band)
and 95\,\% confidence level upper limit calculations. 
  \label{romano_nls1:table:uls}} 
  \begin{tabular}{lcccrr ccc c} 
 \hline 
 \noalign{\smallskip} 
   Source Name                                 & CTA & Expo. &  Energy &  Det.\ c.l.$^{\mathrm{a}}$    & Det.\ c.l.$^{\mathrm{a}}$ & $\overline{TS_{\rm sim}}$  &E$^2$Flux$^{\mathrm{b}}$    & UL             \\   
                                                         & Site &           & Range   &  (TS>10)                            & (TS>25)                         & & $\times10^{-13}$  & $\times10^{-13}$    \\
                                                         &       & (hr)     & (GeV)    & (\%)                                   &   (\%)                              &            &(erg\,cm$^{-2}$\,s$^{-1}$)   &(erg\,cm$^{-2}$\,s$^{-1}$)     \\
 \noalign{\smallskip} 
 \hline 
 \noalign{\smallskip} 
J0324$+$3410                        &N 	   & 50 &   20--150 & 0.0 &    0.0 &  $1.7\pm1.9$   &   $0.89\pm0.97$   &   $ <2.6$  \\
J0932$+$5306                        &N 	   & 50 &   20--150 & 0.0 &    0.0 &  $2.0\pm2.1$   &   $0.91\pm0.97$   &   $ <0.44$  \\
J0937$+$5008                        &N 	   & 50 &   20--150 & 0.0 &    0.0 &  $1.9\pm2.0$   &   $0.88\pm0.94$  &   $ <1.3$  \\
J0958$+$3224                        &N 	   & 50 &   20--150 & 0.0 &    0.0 &  $1.8\pm2.0$   &   $0.84\pm0.96 $ &    $ <0.029$  \\
J1102$+$2239                         &N     & 50 &   20--150 &   0.0 &    0.0 &  $2.0\pm2.0 $  &   $0.93\pm0.98 $ &   $ <2.6$  \\
J1222$+$0413                         &N     & 50 &   20--150 &   0.0 &    0.0 &  $1.8\pm2.0 $  &   $0.84\pm0.95 $  &   $ <1.3$ \\
                                                 &S     & 50 &   20--150 &   1.9 &    0.0 &  $2.3\pm2.2 $  &   $0.76\pm0.77 $  &   $ <0.91$  \\
J1246$+$0238                          &N     & 50 &   20--150 &   0.0 &    0.0 &  $1.8\pm2.0 $  &   $0.90\pm0.96 $   &   $ <2.6$  \\
                                                 &S     & 50 &   20--150 &    2.0 &    0.0 &  $2.2\pm2.2 $  &   $0.76\pm0.76 $  &   $ <3.3$  \\
J1305$+$5116                         &N 	   & 50 &   20--150  & 0.0  &    0.0 &  $2.0\pm2.0$  &   $0.91\pm0.96$  &   $ < 0.32$  \\
J1331$+$3030                         &N      & 50 &   20--150 &   0.0 &    0.0 &  $1.8\pm2.0 $  &   $0.86\pm0.96 $  &   $ <2.5$  \\
J1421$+$3855                         &N 	   & 50 &   20--150 &  0.0 &    0.0 &  $1.9\pm2.0$  &   $0.85\pm0.96$  &   $ <0.50$  \\
J1443$+$4725                        &N 	   & 50 &   20--150 &  0.0 &    0.0 &  $2.0\pm2.0$  &   $0.88\pm0.96$ &   $ <0.032$ \\
J1505$+$0326 Quiescence      
                                               &S 	   & 50 &   20--150 &  2.0 &    0.0 &  $2.3\pm2.2$  &   $0.89\pm0.84$ &   $ <5.6$  \\
J1520$+$4209                        &N 	   & 50 &   20--150 &  0.0 &    0.0 &  $1.9\pm2.0$  &   $0.88\pm0.97$   &   $ <1.8$ \\
J1641$+$3454                        &N 	   & 50 &   20--150 &   0.0 &    0.0 &  $1.9\pm 2.0 $  &   $0.94\pm 0.98 $ & $<0.34$ \\   
J1644$+$2619 Flare                &N 	   & 10 &   20--150 &  34.1 &   0.0 &  $8.8\pm5.5 $  &   $8.35\pm4.41 $&   $<1.6 $  \\
\hspace{47pt}High                  &N  	   & 50 &   20--150 &   0.0 &    0.0 &  $2.2\pm2.2$  &   $1.14\pm1.14 $&   $<0.70 $  \\
\hspace{47pt}Quiescence                       &N  	   & 50 &   20--150 &  0.0 &    0.0 &  $2.0\pm2.0$  &   $0.88\pm0.95 $  &   $ <2.3$  \\
J2007$-$4434                        &S 	   & 50 &   20--150 &  1.8 &    0.0 &  $2.1\pm2.2$  &   $0.74\pm0.76$&   $ <4.8$  \\
J2118$+$0013                       
                                               &S        & 50 &   20--150 &  2.0 &    0.0 &  $2.2\pm2.2$  &   $0.78\pm0.76$ &   $ <5.1 $ \\
J2118$-$0732                          
                                                &S 	   & 50 &   20--150 &  2.2 &    0.0 &  $2.0\pm2.2$  &   $0.74\pm0.75$  &   $ <2.2$  \\
 \noalign{\smallskip}
  \hline
  \end{tabular}
\end{center}
  \end{table*} 
\setcounter{table}{4} 
 \begin{table*} 
\small
 \tabcolsep 4pt  
 \begin{center} 
 \caption{
Results of the simulations of \sbs, \pmn, and \pks{}  with an input model that did not include the cut-off 
due to internal absorption (see Sect.~\ref{romano_nls1:discussion}).  
  \label{romano_nls1:table:dets2}} 
  \begin{tabular}{lcccrr ccc c l } 
  \hline 
 \noalign{\smallskip} 
   Source Name                                 & CTA & Expo. &  Energy &  Det.\ c.l.$^{\mathrm{a}}$    & Det.\ c.l.$^{\mathrm{a}}$ & $\overline{TS_{\rm sim}}$  &E$^2$Flux$^{\mathrm{b}}$    & Notes             \\   
                                                         & Site &           & Range   &  (TS>10)                            & (TS>25)                         & & $\times10^{-13}$\\
                                                         &       & (hr)     & (GeV)    & (\%)                                   &   (\%)                              &            &(erg\,cm$^{-2}$\,s$^{-1}$)      \\
 \noalign{\smallskip} 
 \hline 
 \noalign{\smallskip} 

J0849$+$5108                                  &N  & 10 &    20--30   &    100.0 &  100.0 &  $  76.5\pm 17.0 $  &   $    99.9\pm   11.9 $  & Fig.~\ref{romano_nls1:fig:sed_0850_high_nocut10}\\
	\hspace{8pt} High State            &N  & 10  &   30--50   &    100.0 &  100.0 &  $ 299.1\pm 34.7 $  &   $    92.8\pm    5.8 $  \\
         \hspace{8pt}  No Cutoff           &N  & 10  &   50--75   &    100.0 &  100.0 &  $ 610.1\pm 50.4 $  &   $    82.7\pm    3.6 $  \\  
                                                          &N  & 10  &   75--100 &    100.0 &  100.0 &  $ 779.4\pm 58.4 $  &   $    71.8\pm    2.8 $  \\
                                                          &N  & 10  & 100--140 &    100.0 &  100.0 &  $ 1133.8\pm 72.0 $  &   $    57.0\pm    1.9 $  \\ 
                                                          &N  & 10  & 140--200 &    100.0 &  100.0 &  $ 1035.7\pm 70.7 $  &   $    35.9\pm    1.4 $  \\ 
                                                          &N  & 10  & 200--300 &    100.0 &  100.0 &  $ 431.4\pm 46.0 $  &   $    14.5\pm    0.9 $  \\
                                                          &N  & 10  & 300--400 &    100.0 &   98.7 &  $  51.4\pm 14.5 $  &   $     4.1\pm    0.7 $  \\
\noalign{\smallskip} 

 J0849$+$5108                                 &N  & 50  &    20--30   &   100.0 &  100.0 &  $  75.4\pm 16.6 $  &   $    99.8\pm   11.8 $  & Fig.~\ref{romano_nls1:fig:sed_0850_high_nocut50}\\
	\hspace{8pt} High State            &N  & 50  &    30--50   &   100.0 &  100.0 &  $ 293.9\pm 33.5 $  &   $    92.8\pm    5.7 $  \\
         \hspace{8pt}  No Cutoff           &N  & 50  &    50--75   &   100.0 &  100.0 &  $ 605.8\pm 50.2 $  &   $    82.7\pm    3.7 $  \\
                                                          &N  & 50  &   75--100 &   100.0 &  100.0 &  $ 770.5\pm 59.0 $  &   $    71.7\pm    2.9 $  \\
                                                          &N  & 50  & 100--140 &   100.0 &  100.0 &  $ 1111.4\pm 74.4 $  &   $    57.0\pm    2.1 $  \\
                                                          &N  & 50  & 140--200 &   100.0 &  100.0 &  $ 1044.6\pm 71.7 $  &   $    35.9\pm    1.4 $  \\
                                                          &N  & 50  & 200--300 &   100.0 &  100.0 &  $ 447.3\pm 48.6 $  &   $    14.6\pm    1.0 $  \\
                                                          &N  & 50  & 300--400 &   100.0 &    98.6 &  $  53.2\pm 14.9 $  &   $     4.2\pm    0.7 $  \\
\noalign{\smallskip} 
      J0849$+$5108 Quiescence       &N 	   & 100 &   20--150  & 35.2 &    1.4 &  $   9.0\pm  5.7 $   &   $ <4.1 $   \\                 
      \hspace{8pt}  No Cutoff             &N 	   & 100 &   20--50 &    13.3 &    0.0 &  $   5.4\pm  4.2 $ &   $ <9.2$  \\ 
\noalign{\smallskip} 
J0948$+$0022                                              &N  & 3  &   20--30    &100.0 &   99.5 &  $  58.9\pm 15.7 $  &   $   365.1\pm   52.4 $  & Fig.~\ref{romano_nls1:fig:sed_0948_flare_nocut3} \\
	\hspace{8pt}``Flare'' State                     &S   & 3  &   20--30   &100.0 &  100.0 &  $  67.1\pm 16.5 $  &   $   364.4\pm   49.0 $  \\
         \hspace{8pt}  No Cutoff                        &N  & 3  &   30--50   &100.0 &  100.0 &  $ 149.0\pm 25.3 $  &   $   277.4\pm   25.0 $  \\
							               &S   & 3  &   30--50   &100.0 &  100.0 &  $ 172.0\pm 28.0 $  &   $   277.4\pm   23.9 $  \\
                                                                       &N  & 3  &   50--75   &100.0 &  100.0 &  $ 198.1\pm 30.1 $  &   $   201.0\pm   16.4 $  \\
 							               &S   & 3  &   50--75   &100.0 &  100.0 &  $ 246.6\pm 33.9 $  &   $   201.0\pm   14.9 $  \\
                                                                       &N  & 3  &   75--100 &100.0 &  100.0 &  $ 187.1\pm 29.7 $  &   $   150.0\pm   13.1 $  \\
							               &S   & 3  &   75--100 &100.0 &  100.0 &  $ 228.4\pm 33.1 $  &   $   150.0\pm   12.1 $  \\
                                                                       &N  & 3  & 100--140  &100.0 &  100.0 &  $ 204.8\pm 32.5 $  &   $   103.2\pm    9.3 $  \\
 							               &S   & 3  & 100--140 &100.0 &  100.0 &  $ 261.9\pm 36.8 $  &   $   103.5\pm    8.3 $  \\
                                                                       &N  & 3  & 140--200 &100.0 &  100.0 &  $ 140.1\pm 27.2 $  &   $    55.3\pm    6.4 $  \\
 							               &S   & 3  & 140--200 &100.0 &  100.0 &  $ 186.0\pm 33.4 $  &   $    55.5\pm    6.0 $  \\
                                                                       &N  & 3  & 200--300 &100.0 &   94.0 &  $  46.3\pm 15.4 $  &   $    18.8\pm    4.1 $  \\
 							               &S   & 3  & 200--300 &100.0 &   99.8 &  $  68.4\pm 17.7 $  &   $    18.7\pm    3.2 $  \\
                                                                       &N  & 3  & 300--400 & 19.9  &    0.7 &  $   5.5\pm 5.1$    &   $<1.3$      \\  
 							               &S   & 3  & 300--400 & 33.8  &    1.5 &  $   8.0 \pm 6.2$$^{\rm c}$  &   $4.7\pm2.4$  \\ 
  \noalign{\smallskip} 
                                                        &N  & 5  &   20--150 &   100.0 &  100.0 &  $1362.3\pm 80.6 $  &   $207.3\pm    6.7 $  \\
	                                                &S   & 5 &   20--150  &   100.0 &  100.0 &  $1677.6\pm 87.6 $  &   $207.3\pm    6.1 $  \\
                                                        &N  & 5  &   20--30   &   100.0 &  100.0 &  $96.5\pm 19.8 $        &   $366.1\pm   40.0 $  \\
 							&S   & 5  &   20--30   &   100.0 &  100.0 &  $110.7\pm 21.0 $      &   $366.1\pm   37.0 $  \\
                                                        &N  & 5  &   30--50   &   100.0 &  100.0 &  $246.2\pm 32.7 $      &   $277.3\pm   19.7 $  \\
 							&S   & 5  &   30--50   &   100.0 &  100.0 &  $285.6\pm 36.1 $      &   $277.7\pm   18.6 $  \\
                                                        &N  & 5  &   50--150 &   100.0 &  100.0 &  $1026.4\pm 69.0 $  &   $143.1\pm    5.3 $  \\
 							&S   & 5 &   50--150  &   100.0 &  100.0 &  $1288.1 \pm 83.7 $ &   $142.9\pm    4.8 $  \\
\noalign{\smallskip} 
\noalign{\smallskip} 
J0948$+$0022                   &N  & 5 &    20--30   &  60.8 &    7.6 &  $  13.1\pm  7.2 $  &   $   117.6\pm   40.2 $  & Fig.~\ref{romano_nls1:fig:sed_0948_high_nocut5} \\
	\hspace{8pt} High State           &S   & 5  &   20--30   &  69.6 &    9.2 &  $  14.6\pm  7.5 $  &   $   117.2\pm   37.9 $  \\
         \hspace{8pt}  No Cutoff          &N  & 5  &   30--50   &  98.8 &   66.3 &  $  30.6\pm 11.0 $  &   $    91.1\pm   18.8 $  \\
 							&S   & 5  &   30--50   &  99.7 &   80.7 &  $  35.7\pm 12.1 $  &   $    91.6\pm   17.4 $  \\
                                                                       &N   & 5  &   50--75   &    99.9&   91.5 &  $  42.3\pm 13.5 $  &   $    66.5\pm   11.7 $  \\
 							               &S   & 5  &   50--75   &   100.0&   98.6 &  $  52.7\pm 14.8 $  &   $    66.4\pm   10.1 $  \\
                                                                       &N  & 5  &   75--100 &   100.0&   88.9 &  $  40.9\pm 13.2 $  &   $    49.4\pm    8.8 $  \\
							               &S   & 5  &   75--100 &   100.0&   97.9 &  $  50.8\pm 14.9 $  &   $    49.5\pm    8.0 $  \\
                                                                       &N  & 5  & 100--140 &    99.9 &   94.1 &  $  45.8\pm 14.2 $  &   $    33.9\pm    5.9 $  \\
 							               &S   & 5  & 100--140 &   100.0&   99.3 &  $  59.9\pm 16.3 $  &   $    34.2\pm    5.2 $  \\
                                                                       &N  & 5  & 140--200 &    98.7 &   73.0 &  $  33.0\pm 12.3 $  &   $    18.2\pm    4.0 $  \\
 							               &S   & 5  & 140--200 &    99.9 &   93.0 &  $  44.5\pm 14.3 $  &   $    18.3\pm    3.4 $  \\
 \noalign{\smallskip}
  \hline
  \end{tabular}
\end{center}
\begin{list}{\it Notes.}{} 
  \item[$^{\mathrm{a}}$]{Significance for the detection is high for $TS \geq 25$, low for $10 \leq TS < 25$; source not 
detected for $TS <10$. }
  \item[$^{\mathrm{b}}$]{Upper limits are calculated for 95\,\% confidence level for all cases where $TS <10$. }
  \item[$^{\mathrm{c}}$]{Tentative detection based on 1000 realisations.  }
 \end{list}   
  \end{table*}      
\clearpage           
   
\setcounter{table}{4} 
 \begin{table*} 
\small
 \tabcolsep 4pt  
 \begin{center} 
 \caption{Continued.} 
  \begin{tabular}{lcccrr ccc c l} 
 \hline 
 \noalign{\smallskip} 
   Source Name                                 & CTA & Expo. &  Energy &  Det.\ c.l.$^{\mathrm{a}}$    & Det.\ c.l.$^{\mathrm{a}}$ & $\overline{TS_{\rm sim}}$  &E$^2$Flux$^{\mathrm{b}}$      & Notes             \\   
                                                         & Site &           & Range   &  (TS>10)                            & (TS>25)                         & & $\times10^{-13}$\\
                                                         &       & (hr)     & (GeV)    & (\%)                                   &   (\%)                              &            &(erg\,cm$^{-2}$\,s$^{-1}$)      \\
 \noalign{\smallskip} 
 \hline 
 \noalign{\smallskip} 
                         
 J0948$+$0022                                                                       &N  & 5  & 200--300 &    53.1 &    5.3  &  $  11.8\pm  7.1 $  &   $     6.2\pm    2.6 $  \\
 \hspace{8pt} High State							               &S   & 5  & 200--300 &    78.9 &   17.7 &  $  17.5\pm  9.0 $  &   $     6.2\pm    2.1 $  \\
  \hspace{8pt} No Cutoff                                                                      
                                                                       &N  & 5  & 300--400 &     4.5  &    0.0  &  $   3.1\pm  3.0 $  &   $<0.72$ \\ 
 							               &S   & 5  & 300--400 &     5.7  &    0.0  &  $   3.6\pm  3.3 $  &   $<0.59$ \\ 
\noalign{\smallskip} 
                                               &N  & 5 &   20--150 &  100.0 &  100.0 &  $171.0\pm 27.0 $  &   $69.1\pm6.1$  \\  
                                               &S   & 5 &   20--150 &  100.0 &  100.0 &  $213.8\pm 29.7 $  &   $69.1\pm5.5$  \\
                                               &N  & 5 &   50--150 &  100.0 &  100.0 &  $132.7\pm 24.6 $  &   $47.7\pm4.7$  \\
 						&S   & 5 &   50--150 &  100.0 &  100.0 &  $169.5\pm 27.8 $  &   $47.6\pm4.2$  \\ 
 \noalign{\smallskip} 
\noalign{\smallskip} 
  J1505$+$0326                                              &N  & 5  &   20--30   & 55.7 &    5.4 &  $    12.2\pm  6.9 $  &   $   112.4\pm   39.9 $  & Fig.~\ref{romano_nls1:fig:sed_1505_high_nocut5} \\
		\hspace{8pt}High State                 &S   & 5 &   20--30    &  64.9 &    8.3 &  $    13.7\pm  7.3 $  &   $   112.7\pm   38.1 $  \\
                 \hspace{8pt}  No Cutoff                 &N  & 5  &   30--50   &  98.5 &   62.2 &  $    29.5\pm 10.8 $  &   $    89.3\pm   18.9 $  \\
							               &S   & 5  &   30--50   &  99.4 &   78.1 &  $    34.5\pm 12.0 $  &   $    89.7\pm   17.7 $  \\
                                                                       &N  & 5  &   50--75   &  99.9 &   93.5 &  $    43.9\pm 13.6 $  &   $    67.8\pm   11.5 $  \\
 							               &S   & 5  &   50--75   & 100.0 &   99.1 &  $    54.8\pm 14.9 $  &   $    67.7\pm   10.1 $  \\
                                                                       &N  & 5  &   75--100 &  100.0 &   94.4 &  $    46.2\pm 14.2 $  &   $    52.8\pm    8.9 $  \\
							               &S   & 5  &   75--100 &  100.0 &   99.3 &  $    57.3\pm 15.8 $  &   $    52.9\pm    8.0 $  \\
                                                                       &N  & 5  & 100--140 &  100.0 &   99.8 &  $    61.8\pm 16.5 $  &   $    40.2\pm    5.9 $  \\
 							               &S   & 5  & 100--140 &  100.0 &  100.0 &  $    80.6\pm 18.9 $  &   $    40.3\pm    5.3 $  \\
                                                                       &N  & 5  & 140--200 &  100.0 &   99.7 &  $    60.2\pm 16.2 $  &   $    26.1\pm    4.0 $  \\
 							               &S   & 5  & 140--200 &  100.0 &  100.0 &  $    80.9\pm 19.9 $  &   $    26.1\pm    3.7 $  \\
                                                                       &N  & 5  & 200--300 &   99.5 &   74.0 &  $    34.3\pm 12.9 $  &   $    12.5\pm    2.9 $  \\
 							               &S   & 5  & 200--300 &   100.0 &   97.1 &  $    52.2\pm 16.2 $  &   $    12.6\pm    2.3 $  \\
                                                                        &N  & 5  & 300--400 &   34.0 &    1.6 &  $     8.5\pm  6.1 $  &   $     4.9\pm    2.4 $  \\
 							               &S   & 5  & 300--400 &   54.7 &    6.5 &  $    12.2\pm  7.5 $  &   $     4.8\pm    2.0 $  \\
\noalign{\smallskip}
  \hline
  \end{tabular}
\end{center}
\begin{list}{\it Notes.}{} 
  \item[$^{\mathrm{a}}$]{Significance for the detection is high for $TS \geq 25$, low for $10 \leq TS < 25$; source not 
detected for $TS <10$. }
  \item[$^{\mathrm{b}}$]{Upper limits are calculated for 95\,\% confidence level for all cases where $TS <10$. }
  \item[$^{\mathrm{c}}$]{Tentative detection based on 1000 realisations.  }
 \end{list}   
  \end{table*}

\clearpage
\appendix
\section{Effect of energy dispersion }  \label{romano_nls1:appendix_edisp} 

In the following we address the effect of the energy dispersion on our conclusions 
by considering one exemplary source, \pmn{} in flare, as simulated in 5\,hr.   
The setup of these new simulations, reported in Table~\ref{romano_nls1:table:sims_a1}, 
is the same as that of the earlier simulations performed for this source, with the exception 
of the application of energy dispersion. 
We note, however,  that these simulations were performed with {\tt ctools}  v.\ 1.5.1 
(as opposed to v.\ 1.4.2 as for the rest of this work) 
which removes any noise in the energy dispersion matrix that degraded the precision 
of the energy dispersion computations in earlier software versions. 

\begin{figure*} 
\vspace{-2.3truecm}
\centerline{
 \hspace{-0.5truecm} 
                  \includegraphics[angle=0,width=18.3cm]{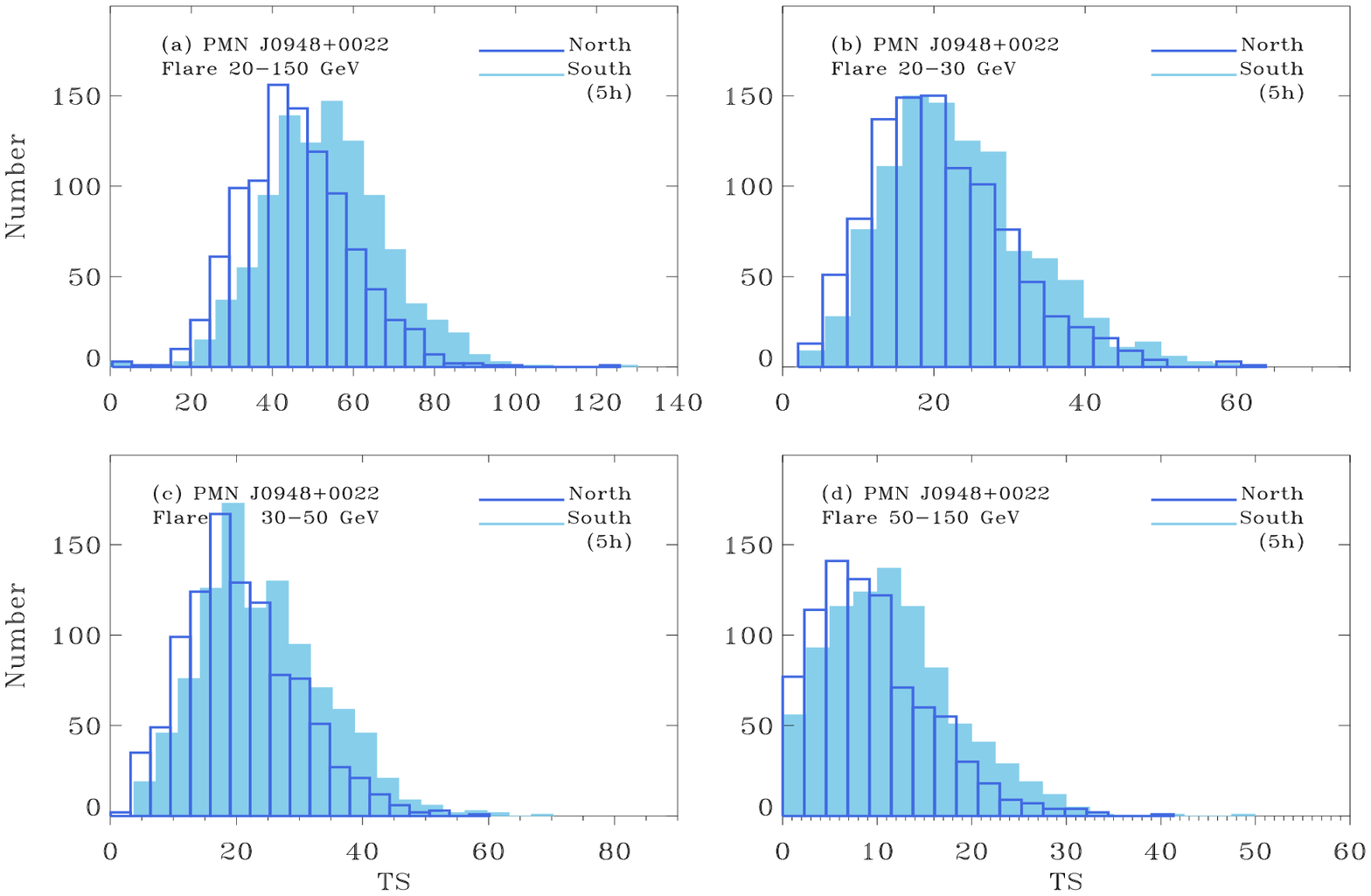}
}
\vspace{-14.5truecm} 
\centerline{
\hspace{-0.5truecm} 
               \includegraphics[angle=0,width=18.3cm]{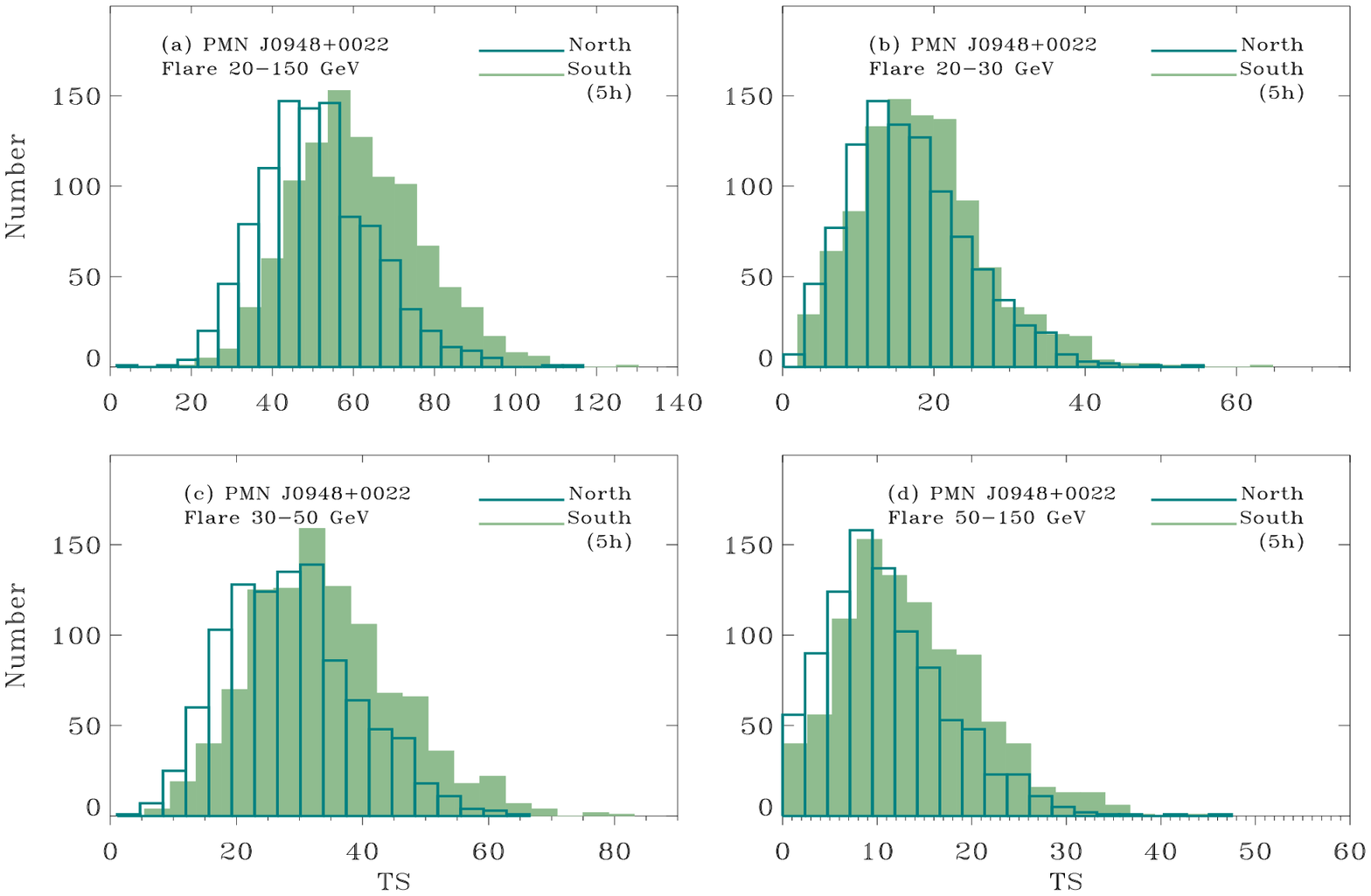}
}
\vspace{-12.truecm} 

\caption{\pmn{} in ``flare'' (exposure 5\,hr): 
comparison of distributions of the TS values depending on the energy band for detection. 
Blue: {\tt edisp=no},  green: {\tt edisp=yes}. 
See Table~\ref{romano_nls1:table:dets} and ~\ref{romano_nls1:table:sims_a1}  for details. }
\label{romano_nls1:fig:dets_over_T_0948_fl5h_cfr} 
\end{figure*}

Figure~\ref{romano_nls1:fig:dets_over_T_0948_fl5h_cfr} shows the comparison of 
the TS distributions for fits performed 
without (blue, top panels) and 
with (green, bottom panels) energy dispersion applied. 
Table~\ref{romano_nls1:table:sims_a2} reports this comparison in terms of 
detection percentages, TS mean values, and derived energy fluxes in each band (Cols.~1--6). 
For ease of comparison, we also report (Col.~7) the corresponding TS mean values for the case 
when the energy dispersion is not applied, as previously reported in Table~\ref{romano_nls1:table:dets}. 

Table~\ref{romano_nls1:table:sims_a2} shows that, 
with the exception of the soft (20--30\,GeV) band,
our approach is a conservative one, 
in that the inclusion of the energy dispersion actually enhances the detection. 
Even for the soft energy band, where these sources are brighter, 
however, the inclusion of the energy dispersion 
does not hamper significantly the detection of the source. 

\newpage

\setcounter{table}{0} 
 \begin{table}
\small 
 \tabcolsep 4pt  
 \begin{center} 
 \caption{Setup of the ({\tt ctools}) simulations to test effects of energy dispersion on  \pmn{} in flare. 
CTA site selected for the simulations: N=North (La Palma), S=South (Paranal).
 \label{romano_nls1:table:sims_a1}} 
  \begin{tabular}{lccccc } 
 \hline 
 \noalign{\smallskip} 
            & Site & IRF                                                       & Expo  & Sim.  & Energy     \\ 
                                         &                             &                                                             & (h)      &  $N_1$  & (GeV)               \\
\noalign{\smallskip} 
 \hline 
 \noalign{\smallskip} 
 &N	&	{\tt North\_z20\_average\_5h}	& 5 &	1000 & 20--150  \\  
&N	&	{\tt North\_z20\_average\_5h}	& 5 &	1000 & 20--30 \\  
&N	&	{\tt North\_z20\_average\_5h}	& 5 &	1000 & 30--50 \\  
&N	&	{\tt North\_z20\_average\_5h}	& 5 &	1000 & 50--150  \\  
&S	&	{\tt South\_z20\_average\_5h}	& 5 &	1000 & 20--150  \\ 
&S	&	{\tt South\_z20\_average\_5h}	& 5 &	1000 & 20--30  \\  
&S	&	{\tt South\_z20\_average\_5h}	& 5 &	1000 & 30--50  \\  
&S	&	{\tt South\_z20\_average\_5h}	& 5 &	1000 & 50--150   \\  
  \noalign{\smallskip}
  \hline
  \end{tabular}
\end{center}
  \end{table}

\setcounter{table}{1} 
 \begin{table}
\small
 \tabcolsep 1pt  
 \begin{center} 
 \caption{Results for \pmn{} in flare (5\,hr exposure) when energy dispersion is applied (Cols.~1--6), compared with the
case when no energy dispersion is applied (Col.~7).   
 \label{romano_nls1:table:sims_a2}} 
  \begin{tabular}{lcccccc c c } 
 \hline 
 \noalign{\smallskip} 
               & CTA &  Energy &  Det.\ c.l.                        & Det.\ c.l.                      & $\overline{TS_{\rm sim}}$  &E$^2$Flux           & & $\overline{TS_{\rm sim}}$$^{\mathrm{a}}$ \\  
                 & Site       & Range   &  (TS>10)                     & (TS>25)                      &                           & $\times10^{-13}$             & &  No Energy\\
                 &       & (GeV)     & (\%)                                 &   (\%)                            &                           &(erg\,cm$^{-2}$\,s$^{-1}$)  & & Dispersion\\
\noalign{\smallskip} 
 \hline 
 \noalign{\smallskip} 
&N      & 20--150 &   99.9 &   98.2 &  $50.9\pm14.3$ &   $ 29.0\pm   6.1 $     & &  $45.8\pm13.7$ \\  
&S       & 20--150 & 100.0 &   99.7 &  $61.0\pm15.6$ &   $ 28.7\pm 5.3$       & &  $53.4\pm14.9$ \\  
&N      & 20--30   &   80.7 &   15.1 &  $16.8\pm7.9$   &   $ 110.3\pm   56.3 $ & &  $21.0\pm9.3$  \\  
&S       & 20--30   &   85.6 &   19.2 &  $18.5\pm8.5$   &   $112.4\pm   54.2 $  & & $23.6\pm9.8$   \\  
&N      & 30--50   &   98.5 &   62.0 &  $28.6\pm10.4$ &   $69.3\pm   18.6 $    & & $20.8\pm9.0$   \\  
&S       & 30--50   &   99.6 &   77.1 &  $34.1\pm11.7$ &   $ 70.5\pm   17.4 $   & & $24.4\pm10.0$ \\  
&N      & 50--150 &   49.7 &    3.4  &  $11.0\pm6.6$   &   $ 8.0\pm    4.7 $      & &  $9.5\pm6.2$$^{\mathrm{b}}$    \\  
&S       & 50--150 &   65.4 &    6.8  &  $13.7\pm7.3$   &   $ 7.8\pm    4.0 $      & &  $11.6\pm6.8$  \\  
  \noalign{\smallskip}
  \hline
  \end{tabular}
\end{center}
\begin{list}{\it Notes.}{} 
  \item[$^{\mathrm{a}}$]{No energy dispersion applied, see full set in Table~\ref{romano_nls1:table:dets}. }
  \item[$^{\mathrm{b}}$]{Considered an upper limit. }
\end{list}   
  \end{table}

\bsp	
\label{lastpage}
\end{document}